# Fabrication of Bilayer Nanofibers from Poly(xylitol dodecanedioic acid) (PXDDA)/Poly(Ɛ-caprolactone) (PCL)/Gelatin Biological Macromolecules and Surface Modification via Spin Coating of Capsules for Skin Wound Treatment

**Mohaddeseh Sharifi and S.Hajir Bahrami[1]**

*Department of Textile Engineering, Amirkabir University of Technology, Tehran, Iran*

## Abstract

In this study, poly (xylitol dodecanedioic acid) (PXDDA) was synthesized from xylitol and dodecanedioic acid (DDA) monomers in a 1:1 molar ratio with Mw = 4038 g/mol using the polycondensation method. Bilayer nanofibers with optimal morphology and average diameter of 271 ± 70 nm were fabricated using electrospinning in voltage of 15 kV and flow rate of 0.5 ml/h, incorporating 15% (w/w) PXDDA, 15% (w/w) gelatin, and 20% (w/w) poly(Ɛ-caprolactone) (PCL) polymer solutions. This diameter of nanofibers was perfectly aligned with genetic algorithm (GA) in ANN model with a cost value of 0.0054 and $R^2 = 0.99$ rather than RSM model. To enhance the stability of the electrospun nanofibers, glutaraldehyde was employed as a crosslinking agent. Additionally, nitrogen doped activated carbon nanoparticles (N-ACNP) served as carriers for clindamycin, intended for spin coating between the layers of PXDDA/Gel/PCL nanofibers. Results indicated that an increase in capsule concentration enhanced bilayer nanofiber contact angle while swelling percentage progressively increased in optimal point of CD concentration over 12 hours. Furthermore, drug release followed Higuchi kinetics, exhibiting high correlation values for the best-fit model. Biodegradability, antibacterial efficacy, cell culture assessments, and MTT assay demonstrated optimizing drug concentration improved cell attachment and viability compared to the control sample.

**Keywords**: Poly(Xylitol Dodecanedioic Acid) (PXDDA), Bilayer Nanofiber, Wound dressings.

[1] Corresponding author.
Email address: hajirb@aut.ac.ir (S.H. Bahrami)

## 1. Introduction

The skin is the largest organ in the human body, composed of multiple layers, including the epidermis, dermis, and hypodermis. While the epidermis serves as the outermost protective layer, shielding the skin from chemical, physical, and biological damage, the dermis, as the second layer, plays a vital role in improving skin function. Through cellular interactions, the extracellular matrix (ECM) forms a dynamic microenvironment [1,2]. Damage to skin tissue, such as tears, burns, and abrasions, poses a significant clinical challenge. Such injuries can lead to bacterial invasion, colonization, and biofilm formation, ultimately resulting in wound infection and, in severe cases, death [3-5]. Wound healing is a complex process involving various immune and structural cells that secrete cytokines, chemokines, and growth factors to repair damaged skin [6-8]. Although wound healing follows four distinct stages, hemostasis, inflammation, proliferation, and remodeling, bacterial infections can prolong the inflammation phase, cause excessive reactive oxygen species (ROS) deposition, accelerate fibroblast aging, hinder angiogenesis, and delay timely treatment, ultimately leading to chronic wounds [8,9]. In other words, disruption in one or more healing stages prevents normal wound healing progression [10,11]. To treat chronic wounds, encapsulating antibiotics, anti-inflammatory drugs, or growth factors in wound dressings may inactivate bioactive drugs, leading to drug resistance and toxic side effects. Various biomaterials, including quantum dots, carbon nanotubes, dendrimers, nanoparticles, nanofibers, porous foams, membranes, and multifunctional hydrogels, have been developed for wound management [12,13]. Ideal wound dressings for biological applications should be easy to design, , preferably be derived from natural sources, absorb and eliminate excessive wound exudate, antibacterial, biocompatible, biodegradable, and non-toxic [13,14]. Electrospun wound dressings offer advantages over conventional dressings, such as high surface area, high porosity for air exchange and effective protection against infection and dehydration, can replicate the ECM structure for skin repair and drug delivery [14,15].

To optimize nanofibers and composites under varying environmental conditions, advanced methodologies such as response surface methodology (RSM) and artificial neural network (ANN) models are employed [16]. RSM is widely used for process modeling and optimization, involving a set of mathematical and statistical tools to analyze and optimize processes by evaluating the effects of independent variables. One key advantage of RSM is its ability to reduce the number of experiments required for statistical analysis of complex processes [14,17]. In addition to RSM, ANN models have emerged as powerful tools for modeling various processes, including electrospinning and composite fabrication. Inspired by biological neural networks, ANNs are constructed with multiple layers of artificial neurons. ANN models are highly effective in capturing complex nonlinear interactions between design variables and performance metrics [17,18].

A multi-layered scaffold can mimic the ECM in skin tissue, accelerate wound healing, protect against contamination and infection, and enable the sustained release of bioactive substances more effectively than single-layer nanofibers. Studies have demonstrated the successful development of bilayer scaffolds for improved wound healing. Asadi et al. fabricated a novel bilayer scaffold with PCL nanofibers reinforced with tannic acid (TA), methacrylate gelatin (GM)/alginate (Al) hydrogel to enhance wound closure rates, effective collagen deposition, and re-epithelialization [19]. Emami et al. developed a bilayer wound dressing incorporating type I collagen nanofibers as the inner layer and collagen/PLLA/Zataria multiflora essential oil nanofibers as the outer layer, exhibiting strong antibacterial, antioxidant, and water absorption properties for improved skin wound healing [20]. Similarly, Kuddushi et al. developed a bilayer membrane with PCL as the top layer, protecting wounds from environmental factors such as moisture and contaminants, while ethylene vinyl alcohol (EVOH) served as the hydrophilic bottom layer, absorbing wound exudates, preventing bacterial infections, and facilitating the healing process. Additionally, incorporating sodium alginate (NaAlg) or zinc oxide (ZnO) into wound dressings proved effective in preventing bacterial infections [21].

Polyesters composed of di-acid and di-alcohol polymers, such as Poly (lactic acid) (PLA), Poly (Ɛ-caprolactone) (PCL), and Poly (Glycerol Sebacate) (PGS), are among the most widely synthesized materials for soft tissue engineering. These materials, approved by the Food and Drug Administration (FDA), possess suitable physical and mechanical properties that mimic the extracellular matrix (ECM). However, their synthesis often involves high complexity, cost, toxic catalysts, large amounts of organic solvents, and reliance on non-renewable resources [5]. To overcome these limitations, researchers have developed Poly (xylitol dodecanedioic acid) (PXDDA), a novel polyester synthesized from renewable resources such as xylitol and dodecanedioic acid (DDA). PXDDA exhibits higher elasticity, lower cost, and favorable mechanical properties while eliminating the need for toxic solvents or excessive catalyst amounts. This makes PXDDA particularly suitable for soft tissue applications in tendons, heart valves, skin, wound dressings, nerves, and blood vessels for drug delivery and tissue regeneration. Xylitol is a kind of carbohydrate that is naturally observed in many vegetables, fruits, corn fibers, and birch wood as a sugar alcohol that is cheap, abundant, and non-toxic with suitable properties including healing, antibacterial, and biocompatibility. Because its solubility in water isn't suitable for supporting cell growth of scaffold for tissue engineering, therefore, a nontoxic cross-linker such as DDA is used to synthesize PXDDA polymer. DDA is dicarboxylic acid to keep blood sugar and energy levels normally, without increasing blood glucose load. [22-26].

Carbon nanoparticles (CNPs) possess unique properties such as large surface area, biocompatibility, nanoscale drug delivery potential, and strong biomolecular binding capacity. These characteristics have made CNPs valuable for adsorption, composite fabrication, and drug delivery applications [27]. Ranjbar et al. incorporated carbon nanoparticles and Au nanoparticles into cellulose nanofibers to detect Staphylococcus aureus (S. aureus) in human blood serum. The nanocomposite, characterized by high surface area, excellent conductivity, and superior biocompatibility, enabled accurate bacterial detection [28]. Additionally, Matteo et al. utilized CNPs as a carrier for copper-based antibacterial agents embedded in PLA nanofibers for food packaging applications [29].

The main goal of this study was to synthesize of PXDDA from its monomers and fabricate a bilayer of nanofibers from PXDDA/Gel/PCL polymer solutions by electrospinning method to treat chronic wounds. To fabricate electrospun nanofibers, interaction between polymer solution concentration, voltage, and flow rate, while minimizing required experiments, was studied using RSM and ANN models. Nitrogen doped activated carbon nanoparticles (N-ACNP) were utilized as carriers for loading clindamycin drugs (CD) into capsules designed for skin wound treatment. To increase the strength and reduce the hydrophilicity of the nanofibers, glutaraldehyde and HCl vapor were employed for crosslinking, conducted in an oven at 35°C for 2.5 hours. Additionally, electrospinning and crosslinking of nanofibers on to the capsule layer enabled controlled drug release, improving skin wound healing outcomes. The fabricated nanofibers were comprehensively evaluated for various properties, including morphology, diameter, hydrophilicity, drug release kinetics, biodegradability, antibacterial properties, cell culture compatibility, and cytotoxicity. These assessments were conducted to determine the effectiveness of the nanofibers in treating chronic wounds.

## 2. Experimental

### 2.1. Materials

Xylitol and Dodecanedioic acid monomers (high-purity) were prepared from Sigma Aldrich. Poly (Ɛ- caprolactone) (PCL) polymer with molecular weights of 80000 Da, carboxymethyl cellulose (CMC) with an average molecular weight of 250,000, urea, Hydrochloric acid (HCL), and glutaraldehyde (GA) were purchased from Sigma Aldrich chemical company. Gelatin powder, ethanol, acetic acid, and formic acid were prepared from Merck Company (Germany). Phosphate buffer saline (PBS, PH= 7.4) were prepared from Roman Industrial Company (USA). Also, clindamycin drugs were purchased from Mofid Pharmaceutical Co.

### 2.2. Methods

#### 2.2.1. Synthesis of Poly (xylitol-co-dodecanedioic acid) (PXDDA) Polymer

Xylitol and dodecanedioic acid monomers (1:1) molar ratio were introduced into a chemical reactor under an $N_2$ atmosphere for polycondensation. The materials were melted at 160°C for 30 minutes. Subsequently, the temperature was reduced to 140°C, and the solution was stirred uniformly for 4 hours under vacuum to increase viscosity. The polymer was then cooled, the vacuum was released, and the final PXDDA polymer was collected and stored [24].

#### 2.2.2. Preparation and crosslinking of PXDDA/Gel/PCL Nanofibers

PXDDA polymer solutions (15%, 20%, and 25% w/w) were prepared using ethanol as a solvent and stirred for 1 hour. However, due to the inability of PXDDA to be electrospun alone, a gelatin polymer solution (15% w/w in acetic acid) was incorporated. The PXDDA/Gel molar ratio was maintained at (1:1) to ensure suitable viscosity for electrospinning. To fabricate mechanically stable nanofibers with a reduced diameter and optimized morphology, PCL polymer solution (20% w/w) dissolved in acetic acid/formic acid (1:1) molar ratio was added to the PXDDA/Gel solution, forming a PXDDA/Gel/PCL blend with a (1:1:1) molar ratio. The blended solution was transferred into a 5 mL syringe for electrospinning using a single-nozzle electrospinning unit.

The electrospinning process was optimized to produce nanofibers with minimal diameter and smooth surfaces. Constant parameters included PCL (20%) and gelatin (15%) concentrations, while PXDDA concentration varied between 15–25%. The optimized electrospinning conditions were: voltage range of 15–25 kV, flow rate of 0.5–1 mL/h, needle-to-collector distance of 18 cm, and a rotational speed of 1000 rpm. Nanofibers were collected on aluminum foil and dried at room temperature. Using Response Surface Methodology (RSM) and Artificial Neural Network (ANN) models, nanofiber diameter and morphology were optimized.

To reduce hydrophilicity and stabilize water absorption, nanofibers were crosslinked via vapor-phase treatment using glutaraldehyde (GA) and hydrochloric acid (HCl) in a (9:1) molar ratio. The crosslinking process was conducted at 35°C for durations of 0.5, 1, 1.5, 2, 2.5, 3, and 4 hours. After crosslinking, the samples were immersed in an aqueous glycine solution (0.2 M) and phosphate-buffered saline (PBS) to neutralize unreacted agents. The samples were then dried overnight in a vacuum oven at 25°C. Scanning electron microscopy (SEM) imaging was used to evaluate the impact of crosslinking duration on nanofiber structure.

### 2.2.3. Surface Modification of PXDDA/Gel/PCL Nanofibers

To enhance antibacterial properties, improve cell attachment, and reduce toxicity, ultimately promoting skin wound healing, clindamycin (CD) was used as an antibiotic agent. To regulate the release of CD from the nanofibers, N-ACNP was employed as a carrier. According to the previous study, N-ACNP was synthesized from carboxyl methyl cellulose and urea by one-pot hydrothermal method [30,31]. N-ACNP, characterized by suitable porosity about 31.8 nm and a negative charge, was combined with positively charged CD at varying concentrations in 1 mL of distilled water. CD was loaded into N-ACNP at concentrations of 0.0005, 0.001, and 0.002 g/mL, while the N-ACNP themselves were maintained at a concentration of 0.001 g/mL. The mixture was stirred for 24 hours to ensure proper encapsulation. Using a spin coating device, the capsules with different concentrations were coated onto the PXDDA/Gel/PCL nanofibers (first layer) at 9000 rpm for 60 seconds, acting as an antibacterial agent. After capsule coating, an additional layer of PXDDA/Gel/PCL nanofibers was electrospun on top (second layer), followed by cross-linking to control the release of CD from the nanofibers. A schematic representation of the wound dressing is shown in Fig. 1.

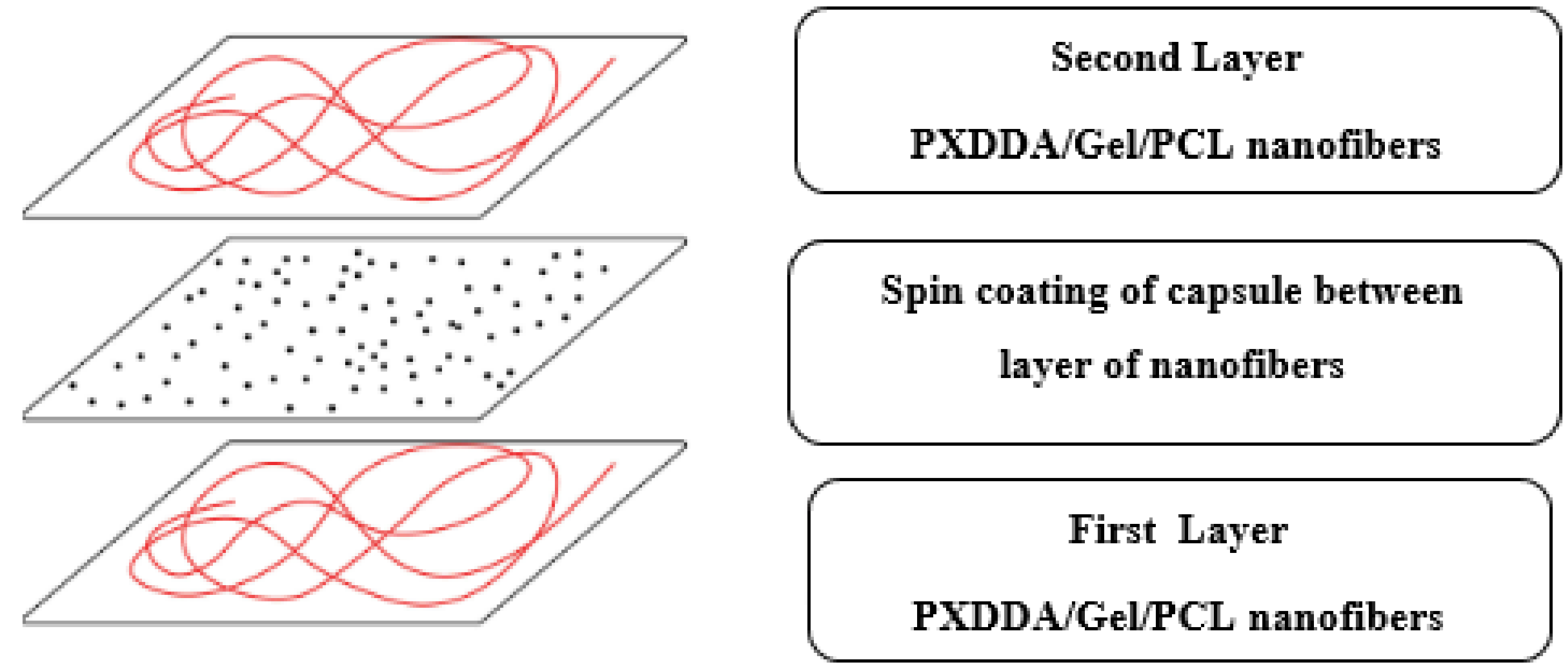

**Fig. 1.** Schematic of the bilayer of wound dressings.

## 2.3. Modeling and Optimization of PXDDA/Gel/PCL nanofibers

### 2.3.1. Response Surface Methodology (RSM) Model

Mathematical and statistical procedures were applied using the Response Surface Methodology (RSM) model to analyze experimental results, identify significant parameters, and optimize the response [14,32-34]. Three key parameters were considered ie, **$X_1$:** Concentration of the PXDDA polymer solution, **$X_2$:** Applied voltage and **$X_3$:** Flow rate (Table 1S).

These parameters influence the diameter of PXDDA/Gel/PCL nanofibers, denoted as **Y** (Eq. (1)) and (Eq. (2)). Additionally, the activation function (**f**) and error parameter (**ε**) were incorporated into the model for improved accuracy.

$$Y = f(X_1, X_2, X_3) \pm \varepsilon \quad (1)$$

$$f = a_0 + \sum_1^3 a_i x_i + \sum_1^3 \sum_1^3 a_{ij} x_i x_j + \sum_1^3 a_{ii} x_i^2 \quad (2)$$

Eq. (2) represents the offset, linear effect of Xi, quadratic effect of Xi, and linear-linear interactions between Xi and Xj. Where $a_0$, $a_i$, $a_{ij}$ and $a_{ij}$ are offset, linear effect of $X_i$, quadratic effect of $X_i$ and linear-linear interactions between $X_i$ and $X_j$, respectively [4,35].

### 2.3.2. Artificial Neural Network (ANN) model

An Artificial Neural Network (ANN) model was employed to predict the optimal input values using a Genetic Algorithm (GA). The GA model searched for optimal values within the input space of the algorithm. A hidden-layer neural network was utilized to predict the nanofiber diameter as output, based on three input parameters defined in (Eqs. (3)- (5)).

In this model X represents the input data matrix and d represents the neural network-predicted diameter. $W^{(1)}$ is the weight matrix of the first layer (3×10 matrix). $b_1$ is the bias matrix of the first layer (1×10 matrix). $W^{(2)}$ is the weight matrix of the second layer (10×1 matrix). $b_2$ is the bias matrix of the second layer (1×1 matrix). The activation functions ($f_1$ and $f_2$) are as follows in supplementary data (Eqs. (1S), (2S)) [35].

$$X = [x_1 \quad x_2 \quad x_3] \tag{3}$$

$$Y_j = f_1(X.W^{(1)}{}_j{}^T + b_{1j}) \quad j = 1 \dots n_H \tag{4}$$

$$d = f_2(Y.W^{(2)^T} + b_2) \tag{5}$$

The general network structure is depicted in Fig. 2. The neural network structure values for $W^{(1)}$, $W^{(2)}$, $b_1$, and $b_2$ were obtained during the learning process [31].

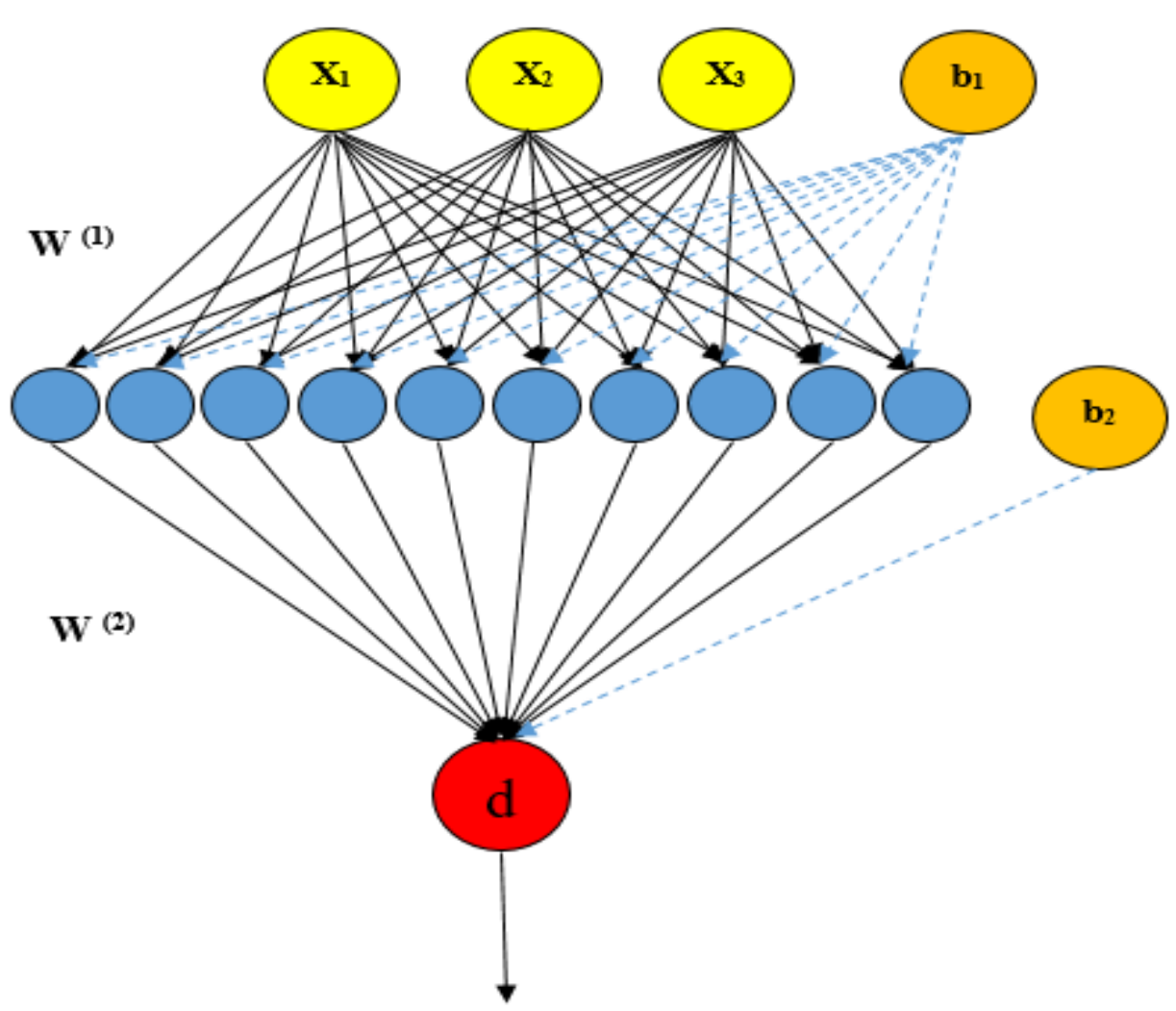

**Fig. 2.** Architecture of developed neural network models with changing concentration of PXDDA polymer solution ($X_1$), voltage ($X_2$), and flow rate ($X_3$) to determine optimal diameter (d) of PXDDA/Gel/PCL nanofibers.

The number of neurons in the hidden layer (nH) was determined experimentally to minimize error, with $R^2$ approaching 1. To achieve this, the Goodness Value (GV) metric was used, balancing both optimization objectives. The (Eqs. (3S) – (7S)) are in supplementary data.

Goodness Value (GV) range was between 0 –2 with an ideal value of 2. Since there are three input variables and one output variable (diameter), the number of input neurons was 3, the number of output neurons was 1 and the number of neurons in the hidden layer was determined experimentally. The activation function of the hidden layer was tan-sigmoid, allowing the network to perform nonlinear modeling. For the output layer, since diameter was being predicted, a linear activation function was selected, ensuring no limitations in approximating output values. Additional network parameters were optimized through experimentation to determine the best Goodness Value (GV) and error minimization. The final network parameters are presented in Table 2S.

#### 2.3.3. Genetic Algorithm (GA)

A Genetic Algorithm (GA) was utilized to determine the optimal data quantity, with parameters detailed in Table 3S. The population type specifies the input format for the fitness function, with a double vector selected when population members are of the double type. The elite count represents the number of individuals guaranteed to advance to the next generation. Reproduction options define the methods by which the GA generates offspring for subsequent generations. The cost limit establishes a threshold for the cost function, terminating the algorithm upon reaching this limit. Additionally, migration, mutation, and crossover serve as key reproduction strategies, enabling the genetic algorithm to generate new individuals for the next generation.

The cost function of the algorithm is as follows (Eq. (6)):

$$minimize : f(\vec{x}) = \sqrt{(d(\vec{x}) - 0.6435)^2} = |d(\vec{x}) - 0.6435| \qquad (6)$$

$$\vec{x}_i = \{x_{1i}, x_{2i}, x_{3i}\} \ \& \ i = 1 \dots n$$

$$0 \leq x_1 \leq 100$$

$$10 \leq x_2 \leq 20$$

$$0 \leq x_3 \leq 1$$

### 2.4. Characterization of PXDDA Polymer

The molecular weight and molecular weight distribution (Mw/Mn) of the PXDDA polymer were determined using Gel Permeation Chromatography (GPC). The polymer was dissolved in tetrahydrofuran (THF) solvent at a concentration of 100 mg/mL and injected into a porous column, where molecules were separated based on their weight and mass. The GPC conditions were set as follows: column oven temperature of 145 °C, mobile phase of 1,2,4-trichlorobenzene (TCB) containing 0.5 g/L butylated hydroxytoluene, with a flow rate of 1 mL/min and injection volume of 0.4 mL.

The fluorescence properties of the PXDDA polymer and its nanofibers were investigated using a QEA0303 fluorescence spectrophotometer at room temperature with an integration time of 5 seconds. The excitation spectra were measured at three different wavelengths ($\lambda$ex = 380, 400, and 420 nm) and scanned using a fluorescence microscope (Nikon Eclipse TE2000-S).

The chemical structure and functional groups of PXDDA were analyzed via Fourier Transform Infrared (FTIR) spectroscopy using a Jasco FT/IR-4100 apparatus over a range of 4000–400 $cm^{-1}$. Additionally, Proton Nuclear Magnetic Resonance ($^1H$ NMR) analysis was conducted using a Bruker Avance 400 MHz spectrometer. The PXDDA polymer was dissolved in acetone at 25 °C, and the data was analyzed using MestReNova software.

To determine the degree of crystallinity ($\chi c$), X-ray diffraction (XRD) analysis was performed using a Siemens D5000 powder X-ray diffractometer. The mass loss and its derivative were assessed using Thermogravimetric Analysis (TGA). TGA analysis was conducted with a Setaram TGA 500 apparatus, applying a heating rate of 10 °C/min for a 20 mg PXDDA polymer sample. The sample was purged in a platinum pan and heated in a nitrogen ($N_2$) atmosphere from 25 °C to 800 °C to evaluate the weight loss percentage and its temperature relationship.

## 2.5. Characterization of Nanofibers

### 2.5.1. Scanning Electron Microscopy (SEM)

The morphology and diameter of nanofibers were characterized using Scanning Electron Microscopy (SEM). Samples were sputter-coated with gold at an accelerating voltage of 15 kV, and images were captured at various magnifications. To determine the mean diameter and standard deviation of 100 nanofibers, statistical analysis was performed using Digimizer and SPSS software.

### 2.5.2. Fourier transform infrared (FTIR)

Fourier transform infrared spectroscopy (FTIR) (Nicolet Nexus 670, USA) was utilized by mixing 1 mg sample in 100 mg KBr to identify the chemical structure, functional groups, and possible interaction between the materials in the wave number of 4000- 400 $cm^{-1}$. These samples include clindamycin drug (CD), N-ACNP, capsule, PXDDA/Gel nanofibers, PCL nanofibers, one layer of PXDDA/Gel/PCL without crosslinked nanofibers, one layer of PXDDA/Gel/PCL crosslinked nanofibers, and bilayer of PXDDA/Gel/PCL crosslinked nanofibers.

### 2.5.3. X-ray Diffraction (XRD) Analysis

X-ray diffraction (XRD) was performed to determine the degree of crystallinity (χc) of N-ACNP, capsule, one layer of PXDDA/Gel/PCL without crosslinked nanofibers, and bilayer nanofibers with coating capsule in different concentrations of CD. The XRD pattern was characterized using a Siemens D5000 powder X-ray diffractometer. The samples were placed in an acrylic sample holder approximately 3 mm deep and analyzed using parallel beam optics with CuKα radiation at 40 kV and 30 mA. The diffraction angle was scanned from 0° to 110°. The crystallinity degree (χc) of the samples was calculated using Eq. (7):

$$\chi_c = \frac{A_c}{A_c + A_a} \quad (7)$$

Where $A_a$ represents the area of amorphous diffraction peaks and $A_e$ represents the area of crystalline diffraction peaks.

### 2.5.4. Surface Hydrophilicity of Nanofibers

The surface hydrophilicity of single-layer and bilayer PXDDA/Gel/PCL nanofibers, both crosslinked and coated with capsules at different concentrations, was evaluated using a contact angle measurement instrument. A water droplet was placed on the nanofiber surface, and images were captured at 1-second and 2-second intervals using a video contact angle system (SSC DC318P color video camera, Sony Co., New York). Contact angles were measured using Digimizer software, and each sample was tested three times for accuracy.

### 2.5.5. Swelling of Nanofibers

The swelling degree of bilayer nanofibers with different capsule concentrations was measured by assessing weight changes upon immersion in phosphate buffer (pH = 7.4) at 37°C for 24 hours. Samples were reweighed ($M_t$) at different time intervals: 30 minutes, 1.5 hours, 3 hours, 6 hours, 12 hours, and 24 hours. The bulk hydrophilicity of nanofibers was calculated using Eq. (8):

$$\Delta w = \frac{M_t - M_i}{M_t} \times 100 \quad (8)$$

Where $M_t$ is the weight at time t, and $M_i$ is the initial weight.

### 2.5.6. Biodegradability of Nanofibers

The degradability of bilayer nanofibers was assessed by immersing them in PBS solution (pH = 7.4). The nanofibers included Control: PXDDA/Gel/PCL nanofibers without modifications, Surface-modified nanofibers: Loaded with CD/N-ACNP in varying concentrations of 0.0005/0.001 g/mL, 0.001/0.001 g/mL and 0.002/0.001 g/mL. The samples were placed in a shaking incubator for 1, 3, 5, 7, 10, and 30 days. Following incubation, they were washed with distilled water, dried at room temperature, and analyzed using SEM to examine morphology changes.

### 2.5.7. In Vitro Drug Release

The release profile of CD within PXDDA/Gel/PCL nanofiber layers was examined in PBS solution (pH = 7.4, at 37°C). CD absorption at various concentrations (0.0001, 0.0005, 0.001, and 0.005 g/mL) was evaluated using a UV spectrophotometer (λ = 190–250 nm), with peak absorption occurring at 203 nm. A calibration curve of UV absorption was plotted against drug concentration, based on the Beer-Lambert equation (Eq.(9)), with an R-value of ~0.9416 (Figure 1S).

$$A = \varepsilon l c \qquad (9)$$

A is the absorption, c is the drug concentration, l is the distance of light passes through the material and Ɛ is the extinction coefficient.

Nanofibers containing different capsule concentrations were incubated in a shaking incubator at 37°C. To maintain a constant volume, 3 mL of PBS solution was removed from each tube at designated intervals and replaced with an equivalent volume of fresh PBS. The amount of drug release at various concentrations was quantified by recording UV-vis absorption spectra at 203 nm. The cumulative drug release from the nanofibers was measured over 72 hours, and the percentage of cumulative drug release ($E_r$) was calculated using Eq. (10).

$$E_r\,(\%) = \frac{V_0 \times C_n + V_r \times \sum_1^{n-1} C_i}{m_{totl}} \times 100 \qquad (10)$$

Where $m_{total}$ is the amount of CD that was encapsulated in the N-ACNP and coated on the surface of nanofibers, $V_0$ and $V_r$ are the volume of the release media and the replace media, respectively. Also, $C_n$ is the drug concentration in the sample [36].

### 2.5.8. Kinetics of clindamycin drug release

There are different models to investigate the release CD. These models are shown in (Table 4S) with their equations including Eqs. (8S)- (11S) are in supplementary data

In these models, $M_t$ exhibits the release drug concentration at the time of t, $M_0$ exhibits the initial drug concentration in solution, and $K_0$, $K_1$, and $K_{HC}$ are the constants of drug release rate for zero-order, first-order, and Higuchi kinetics models. In Korsmeyer-Peppas model $\frac{M_t}{M_\infty}$ is the released drug fraction at the time of t, $K_{kp}$ shows the constant of the release rate, and n demonstrates the release exponent [36].

### 2.5.9. Antibacterial Properties

#### 2.5.9.1. Measurement of Microbial Growth Using Suspension (Quantitative Method)

The antibacterial properties of the samples were evaluated using the AATCC Test Method 100-2004. To prepare a $1.5 \times 10^5$ CFU/mL bacterial suspension, a portion of microorganism colonies was transferred into a test tube containing Tryptic Soy Broth (TSB). The optical density (OD) of the suspension was adjusted to McFarland 0.5 ($10^7$ CFU/mL) with an absorbance range of 0.08–0.10 at 600 nm. Samples were cut into circular discs (48 ± 1 mm in diameter) and sterilized using UV light for 45 minutes. Then, 1 mL of microbial suspension ($10^7$ CFU/mL) prepared in liquid culture medium was added to the samples placed in 100 mL Media-Lab Bottles, and incubated at 37°C for 24 hours to allow bacterial growth.

The antimicrobial activity was evaluated against Staphylococcus aureus (Gram-positive) and Escherichia coli (Gram-negative) bacteria. After 24 hours, 100 mL of physiological saline was added to each bottle and shaken vigorously. Then, 1 mL of the microbial suspension was serially diluted ($10^0$, $10^1$, and $10^2$) and transferred to Petri dishes. Next, 15 mL of melted Tryptic Soy Agar (45°C) was added to each dish. The dishes were gently shaken in a figure-eight motion to ensure thorough mixing. After solidification at 25°C, the Petri dishes were inverted and incubated at 37°C for 24 hours in a natural atmosphere incubator. After incubation, ImageJ software was used to count colony-forming units (CFU/mL) adjacent to the tested nanofibers.

#### 2.5.9.2. Measurement of Microbial Growth Using Inhibition Zone (Qualitative Method)

The antibacterial properties of N-ACNP and bilayer PXDDA/Gel/PCL nanofibers were assessed using capsules with varying CD concentrations (0.0005 mg/mL, 0.001 mg/mL, and 0.002 mg/mL) against S. aureus and E. coli bacteria. Bilayer nanofibers ($2 \times 2$ cm$^2$) were prepared and sterilized under UV light for 2 hours. The samples were incubated in a culture medium at 37°C for 24 hours, and the inhibition zones around the nanofibers were measured.

### 2.5.10. Cell Viability

To assess the cytotoxicity of nanofibers, mesenchymal stem cell (MSC) viability was evaluated using the MTT assay (3-(4,5-dimethylthiazol-2-yl)-2,5-diphenylterazolium bromide test). MSC cultures were prepared using Dulbecco's Modified Eagle Medium (DMEM) with 10% PBS, incubated at 37°C in a 5% $CO_2$ atmosphere. Samples were sterilized using UV light for 2 hours and immersed in PBS for 30 minutes, followed by overnight incubation. Nanofibers were placed in 48-well culture plates containing MSCs ($1 \times 10^4$ cells/500 µL of medium) for 24 hours. The wells were washed with PBS, and 100 µL of MTT solution (0.5 mg/mL in PBS) was added. The samples were incubated at 37°C for 30 minutes, allowing the formation of purple formazan crystals, which indicate viable cells.

After incubation, the medium was discarded, and 200 µL of dimethyl sulfoxide was added to dissolve the formazan crystals. The optical density was measured using an ELISA reader at 570 nm, and cell viability (%) was calculated using Eq. (11) [36].

$$\text{Cell viability (\%)} = \left(\frac{Average\ optical\ density\ of\ sample}{Average\ optical\ density\ of\ control}\right) \times 100 \qquad (11)$$

Measurements were taken over 1, 2, and 3 days.

### 2.5.11. Cell Culture

To evaluate cell attachment and growth, nanofibers (0.5 mm$^2$) were sterilized under UV light for 2 hours. 2,000 MSCs were seeded onto nanofibers placed in 96-well culture plates, with 200 µL of culture medium added to each well. The plates were incubated at 37°C for 1, 3, and 5 days. After 5 days, scaffolds were removed from the culture medium, and cell adhesion was analyzed using SEM.

#### 2.5.12. Statistical Analysis

Statistical analysis was performed using Response Surface Methodology (RSM) and Artificial Neural Network (ANN) models in Design Expert and Python software. In Design Expert, a one-way ANOVA test was conducted to determine the optimal sample, with a P-value $< 0.05$ indicating statistical significance. All quantitative results were reported as mean ± standard deviation (SD). Additionally, population size and cost value were applied in a Genetic Algorithm (GA) model to predict the optimal sample for nanofiber diameter and morphology.

## 3. Results and Discussion

### 3.1. Characterization of PXDDA Polymer

#### 3.1.1. Molecular Weight of PXDDA Polymer

Gel Permeation Chromatography (GPC) is a widely used technique for determining polymer molecular weight, which significantly influences polymer properties. In this method, polymer molecules are separated based on their size to calculate the weight-averaged molecular weight (Mw), number-averaged molecular weight (Mn), and poly dispersity index (PDI). The results indicate a linear calibration curve consisting of 12 data points, as presented in Fig. 3(a). The monomers of this polymer followed the same linear calibration curve under GPC conditions. Additionally, GPC analysis revealed unimodal characteristics, with values of Mw = 4038 g/mol, Mn = 3043 g/mol, and narrow molecular weight distribution (Mw/Mn = 1.32), indicating ultrahigh molecular weight properties, as shown in Fig. 3(b).

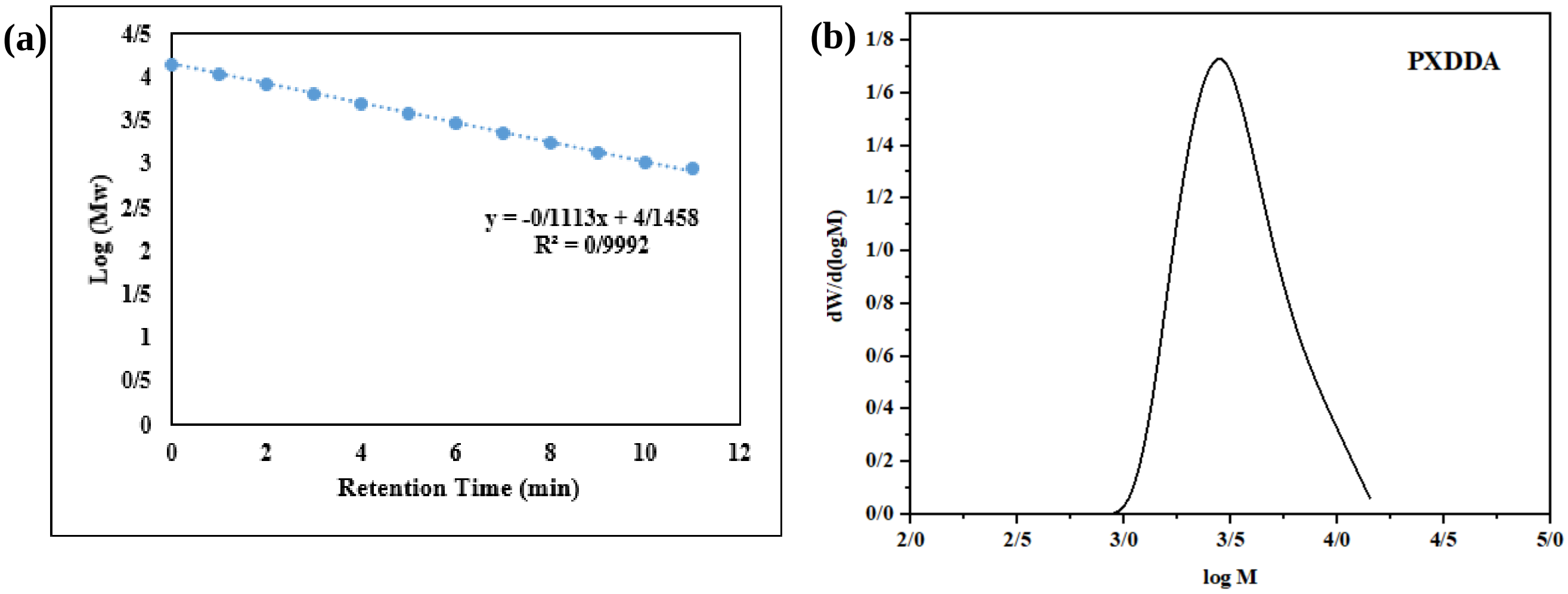

**Fig. 3.** a) Calibration curve prepared with PXDDA polymer standards for GPC with THF solvent, and b) Stacked gel permeation chromatography (GPC) traces of PXDDA polymer.

### 3.1.2. Fourier Transform Infrared (FTIR) Analysis

The chemical structure and functional groups of the PXDDA polymer, synthesized from xylitol and dodecanoic acid, were analyzed using Fourier Transform Infrared (FTIR) spectroscopy, as shown in Fig. 4. The carbonyl (C=O) and hydroxyl (O-H) groups originate from dodecanedioic acid, while the hydroxyl group is the key functional group in xylitol. The interaction between the acidic and alcoholic groups of the monomers leads to the formation of carbonyl groups during polymerization. A sharp peak at 1738 $cm^{-1}$ corresponds to the C=O stretch, while peaks at 1048-1181 $cm^{-1}$ is assigned to the C-O functional groups of xylitol. Peaks at 1392 $cm^{-1}$ and 1463 $cm^{-1}$ correspond to $CH_2$ and $CH_3$ groups, respectively. The C-H groups in alkanes or aromatic compounds appear at 2954 $cm^{-1}$, and the hydroxyl (O-H) groups of xylitol are represented by a peak at 3418 $cm^{-1}$. According to previous studies, FTIR analysis of the PXDDA polymer confirmed peaks at 1735 $cm^{-1}$, 1120 $cm^{-1}$, and 3378 $cm^{-1}$, corresponding to C=O, C-O, and O-H functional groups, respectively [22,24,26].

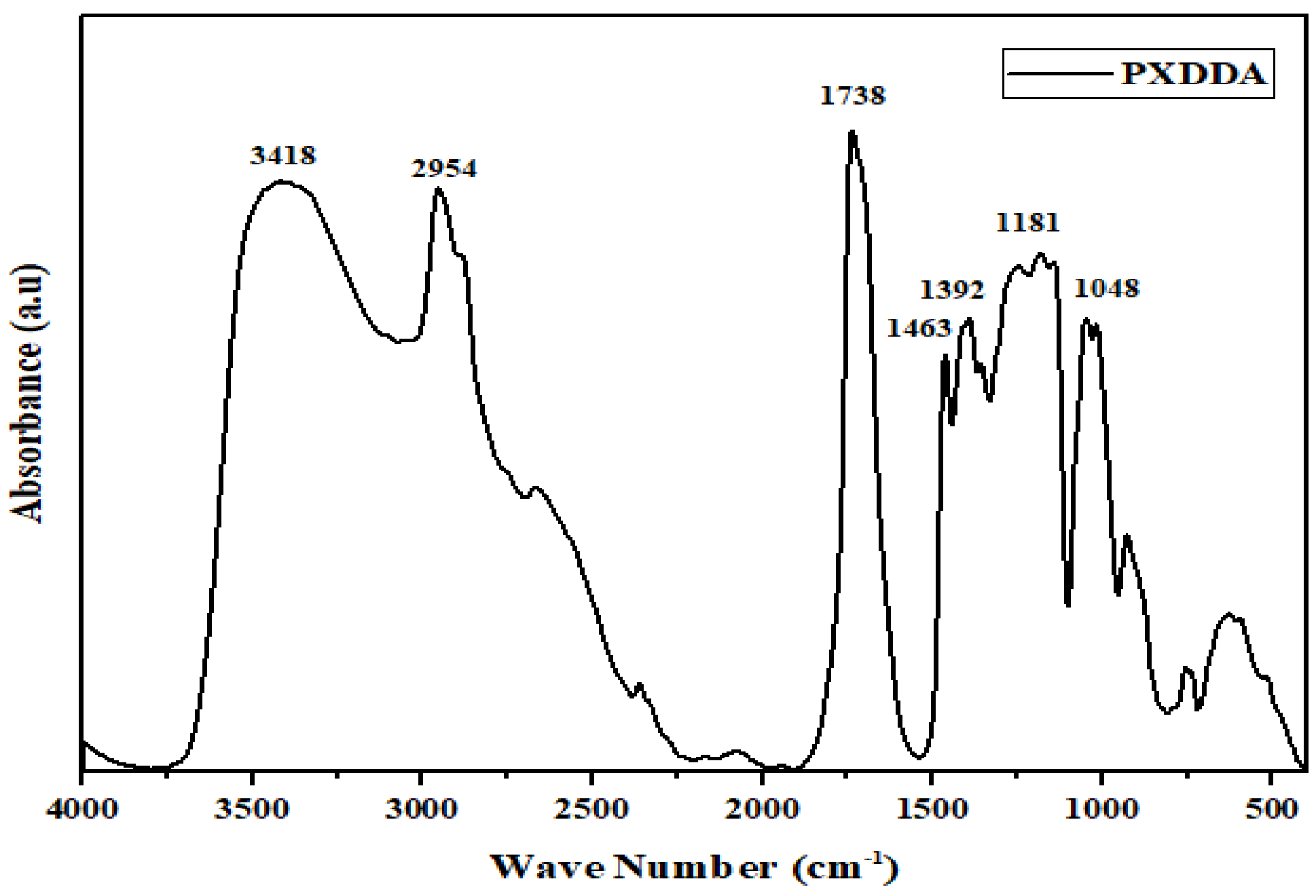

**Fig. 4.** The functional groups of PXDDA polymer in different wave numbers.

### 3.1.3. $^1$H NMR (Nuclear Magnetic Resonance) Analysis

The molecular structure of the synthesized PXDDA polymer was examined using proton Nuclear Magnetic Resonance ($^1$H NMR) spectroscopy, as shown in Fig. 5. The polymer sample was analyzed using MestReNova software, with $^1$H NMR (400 MHz, Acetone-d) presenting different chemical shifts (δ values) 4.22–4.09 ppm (multiplet, 3H), 3.63 ppm (doublet, J = 5.8 Hz, 4H), 2.36 ppm (double triplet, J = 26.4, 6.1 Hz, 5H), 2.10 ppm (doublet, J = 5.0 Hz, 0H) and1.65 ppm (double doublet, J = 7.2, 3.6 Hz, 5H). Aliphatic compounds typically exhibit signals within the 1.2–3.6 ppm range. Peaks at 2.3–2.4 ppm correspond to protons adjacent to COOH groups, while the 1.65 ppm peak is attributed to -$CH_2$ groups in the DDA structure. Peaks within 3.63–4.22 ppm represent all protons of xylitol, with the 3.63 ppm peak corresponding to -$CH_2$ groups near electronegative atoms. Peaks at 4.09–4.22 ppm are associated with -CH groups linked to hydroxyl (OH) functional groups in xylitol units. Previous studies confirmed similar peaks, including 1.25 ppm, attributed to -$CH_2$ groups in the DDA monomer, 1.59 ppm, associated with -$CH_2$ groups near COOH (HOOC-$CH_2$-$CH_2$), 2.2–2.4 ppm, corresponding to protons adjacent to COOH groups, 3.67–4.73 ppm, representing protons of OH groups in xylitol and 3.76 ppm, indicating -$CH_2$ groups interacting with functional groups in xylitol [22,24,26].

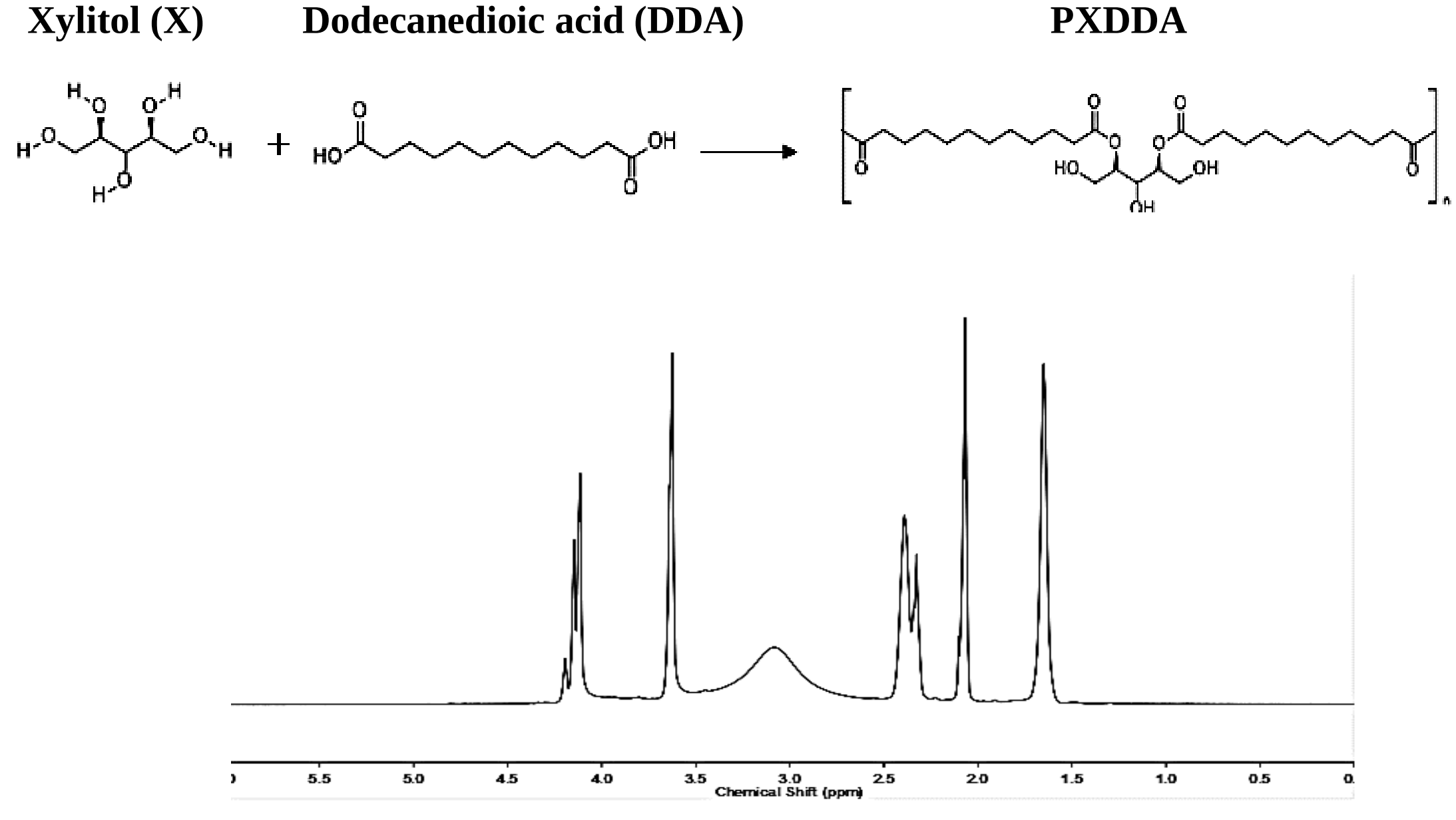

**Fig. 5.** HNMR spectra of PXDDA polymer

#### 3.1.4. X-ray Diffraction (XRD) Analysis

The degree of crystallinity of the PXDDA polymer is shown in Fig. 2S. According to the XRD pattern, nearly all crystalline peaks have disappeared, with only a steamed-bun-like peak observed in the range of 2θ = 15–35°. This suggests that the PXDDA polymer is predominantly amorphous rather than crystalline. The amorphous nature of the PXDDA polymer can be attributed to irregular polymer chains, caused by electrostatic interactions and intermolecular hydrogen bonding between polymer chains. Additionally, the prepolymer was subjected to vacuum conditions to increase solution viscosity, resulting in a gel-like state. Other reports confirmed the amorphous structure of PXDDA polymer in the range of 2θ = 15–25° [23,24,26].

#### 3.1.5. Auto fluorescence of PXDDA Polymer and Its Nanofibers

The characteristic fluorescent properties of polymers are widely utilized in biological research to enable imaging for drug tracing. The PXDDA polymer exhibits auto fluorescence, primarily due to variations in C=O and C–O bonds in its prepolymer and polymer structures. During polymer synthesis, the curing process leads to the formation of C=O linkages, which enhance its fluorescent properties. The fluorescence characteristics of the PXDDA polymer were analyzed across the wavelength range of 420–720 nm. Fig. 3S(a) presents the fluorescence emission spectra of the PXDDA polymer at three excitation wavelengths (380 nm, 400 nm, and 420 nm), represented in blue, red, and black, respectively. According to Firoozi et al., the PXDDA polymer exhibits a maximum emission at ~520 nm, attributed to the presence of specific functional groups in its structure [24]. Similarly, Figure 3S(b) illustrates the fluorescence emission spectra of PXDDA nanofibers, showing a maximum wavelength at 430 nm, along with additional emissions at 480 nm and 520 nm.

### 3.1.6. Thermo gravimetric analysis (TGA)

Thermogravimetric analysis (TGA) was conducted to examine mass changes in the newly synthesized polymer as temperature increased. As shown in Fig. 6, three distinct weight loss stages were observed, corresponding to oxidation, decomposition, and physical processes such as evaporation. In the first stage, the polymer lost approximately 10% of its weight up to 350°C, primarily due to moisture evaporation. The second stage, occurring between 350°C and 460°C, exhibited a sharp weight reduction of about 72%, attributed to polymer degradation and decomposition. In the final stage, from 460°C to 510°C, a gradual weight loss occurred due to the evaporation of polymer degradation products. By 510°C, the remaining polymer weight stabilized at approximately 10%. Similar trends have been reported in previous studies [22,24].

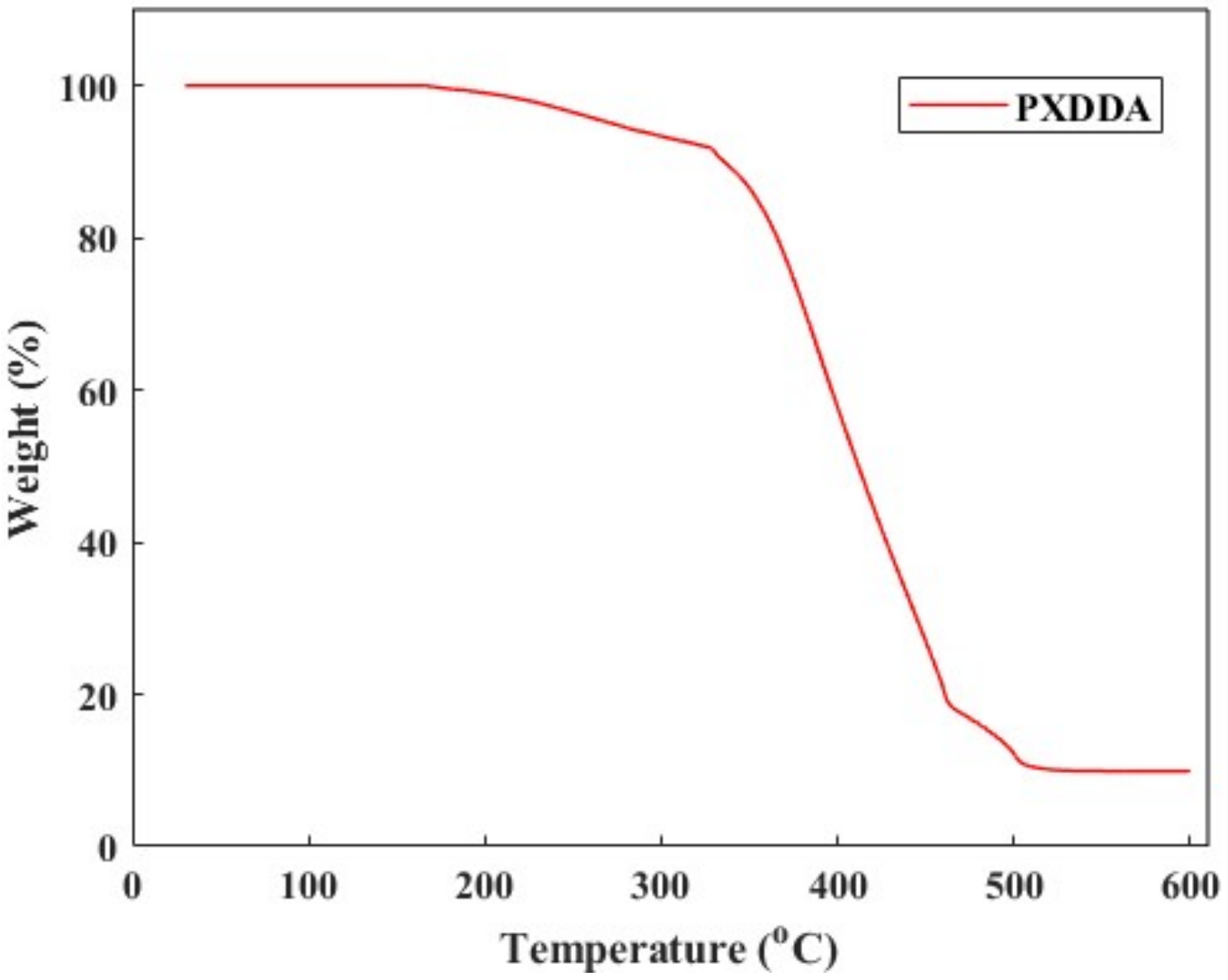

**Fig. 6.** TGA curve of PXDDA polymer.

### 3.2. Modeling and optimization of PXDDA/Gel/PCL Nanofibers

#### 3.2.1. Response Surface Methodology (RSM) Model

To determine the optimal morphology and diameter of nanofibers, different modeling approaches, including Response Surface Methodology (RSM) and Artificial Neural Network (ANN) models, were employed. In the RSM model, a central composite design was utilized, enabling interaction analysis between parameters with fewer experiments to identify optimal samples. Various experimental parameters were varied to investigate their effect on the nanofiber's diameter. These parameters were PXDDA polymer solution concentration: 15–25% (w/w), voltage: 15–25 kV and flow rate: 0.5–1 mL/h. The central composite design, presented in Table 5S, examined the impact of the lowest and highest parameter values on nanofiber diameter. For this PXDDA polymer solution concentrations were 11.5% and 28.4% (w/w). Voltage was 12.9 kV and 27 kV and flow rate: 0.3 mL/h and 1.1 mL/h. For electrospinning PXDDA/Gel/PCL nanofibers, certain parameters such as gelatin polymer concentration: 15% (w/w), PCL polymer concentration: 20% (w/w), polymer solution ratio: PXDDA/Gel/PCL = (1:1:1) and distance between collector and needle tip: 18 cm remained constant. The inclusion of PCL polymer solution in the nanofiber fabrication process resulted in a decrease in nanofiber diameter, attributed to entanglement between polymer chains. Without PCL, increasing the PXDDA polymer concentration from 10% (w/w) to 18% (w/w) led to nanofiber diameter growth from 395 ± 76 nm to 1222 ± 235 nm (Fig. 4S). Thus, PCL polymer solution was incorporated with PXDDA/Gel to fabricate nanofibers. The best nanofiber morphology and diameter were observed in Run 2, achieving a diameter of 271 ± 70 nm, produced at PXDDA polymer concentration: 15% (w/w), gelatin polymer concentration 15% (w/w), PCL polymer concentration 20% (w/w), voltage: 15 kV and flow rate: 0.5 mL/h.

The effect of different parameters on nanofiber diameter was visualized using single-factor impact diagrams (Fig. 5S). Correlation analysis identified the PXDDA polymer solution concentration as the most influential factor on nanofiber diameter. SEM images (Fig. 6S) revealed that increasing PXDDA polymer concentration from 11.5% (w/w) to 28.7% (w/w) resulted in an increase in nanofiber diameter from 211 ± 36 nm to 478 ± 192 nm. This was attributed to a rise in polymer solution viscosity, which directly impacted nanofiber diameter (Fig. 7). The ANOVA table identified parameters with P-values $< 0.05$ as significant factors affecting nanofiber diameter (Table 1). P-value for PXDDA polymer solution concentration was 0.0001 and for voltage it was 0.0006. The RSM model exhibited statistical significance, with a low lack-of-fit value of 0.0069.

Additionally, parameter interactions including polymer concentration, voltage, and flow rate were statistically significant (P ≈ 0.0005 and 0.0463, respectively). The 3D surface plot (Fig. 7S) illustrates the effect of key parameters (polymer solution concentration, voltage, and flow rate) on nanofiber diameter. (Eqs. (12) and (13)) describe relationships between coded and actual factors, highlighting their influence on nanofiber diameter. The RSM model prediction closely aligned with experimental results, with regression analysis presented in Figure 8S. The RSM model predicted the optimal nanofiber sample, achieving a diameter of 351 ± 97 nm could be produced with PXDDA polymer concentration: 20% (w/w), voltage 20 kV, flow rate: 0.7 mL/h with $R^2$ value: 0.98.

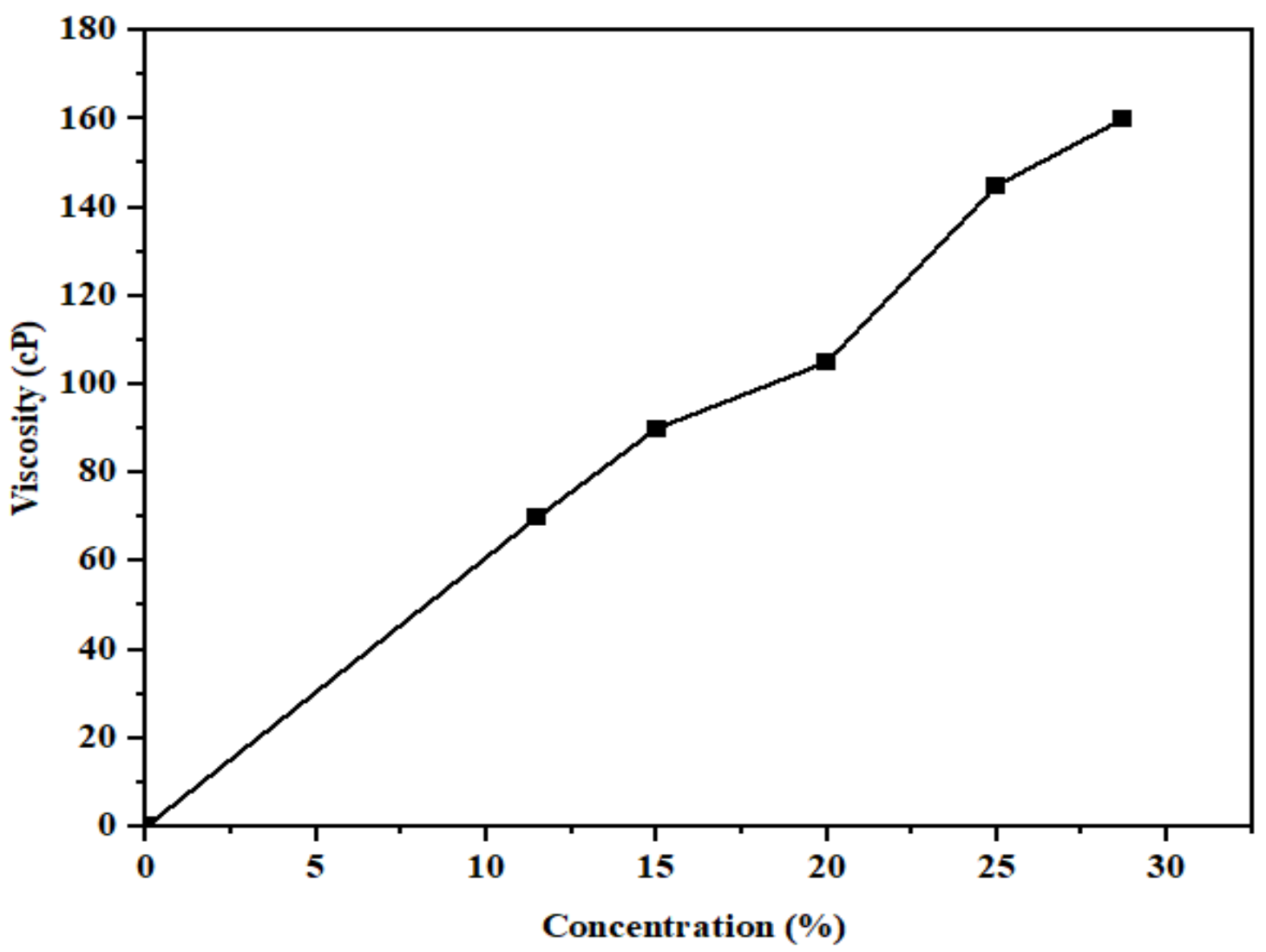

**Fig. 7. The viscosity of PXDDA/Gel/PCL polymer solutions.**

**Table 1.** ANOVA table for statistical analysis on the diameter of nanofibers.

| Source | Sum of Squares | df | Mean Square | F-value | p-value | |
|---|---|---|---|---|---|---|
| Model | 93537.35 | 9 | 10393.04 | 105.94 | 0.0001 | Significant |
| $X_1$-Concentration | 82736.89 | 1 | 82736.89 | 843.33 | 0.0001 | |
| $X_2$-Voltage | 2427.73 | 1 | 2427.73 | 24.75 | 0.0006 | |
| $X_3$-Flow rate | 1.90 | 1 | 1.90 | 0.0194 | 0.8920 | |
| $X_1X_2$ | 2541.13 | 1 | 2541.13 | 25.90 | 0.0005 | |
| $X_1X_3$ | 10.53 | 1 | 10.53 | 0.1074 | 0.7499 | |

| $X_2X_3$ | 506.89 | 1 | 506.89 | 5.17 | 0.0463 | |
|---|---|---|---|---|---|---|
| $X_1^2$ | 295.32 | 1 | 295.32 | 3.01 | 0.1134 | |
| $X_2^2$ | 4985.36 | 1 | 4985.36 | 50.82 | 0.0001 | |
| $X_3^2$ | 7.15 | 1 | 7.15 | 0.0729 | 0.7927 | |
| Residual | 981.07 | 10 | 98.11 | | | |
| Lack of fit | 910.72 | 5 | 182.14 | 12.94 | 0.0069 | Not significant |
| Pure Error | 70.35 | 5 | 14.07 | | | |
| Core Total | 94518.43 | 19 | | | | |

Code equation:

$$77.83X_1 + 14.24\,X_2 – 0.3731\,X_3 +17.82\,X_1X_2 + 1.15\,X_1X_3 +7.96\,X_2X_3 – 4.51\,X_1^2 -23.90\,X_2^2 - 0.7009\,X_3^2 = +355.03 \quad (12)$$

Actual equation

$$7.82869\,X_1 +23.04855\,X_2 – 130.39091\,X_3 + 0.712900\,X_1X_2 + 0.918000\,X_1X_3 + 6.36800\,X_2X_3 – 0.180205\,X_1^2 -0.955872\,X_2^2 – 11.21436\,X_3^2 = -78.43067 \quad (13)$$

### 3.2.2. Artificial Neural Network (ANN) and Genetic Algorithm (GA) Models

The results indicate that the parameters exhibit a complex non-linear relationship. The Response Surface Methodology (RSM) model is linear and lacks reliability in handling such complexity. Consequently, non-linear relationships between variables in the ANN model were employed for improved performance compared to the RSM model. Fig. 9S illustrates the correlation between the experimental and predicted diameters of nanofibers, demonstrating a strong model fit and high generalizability. Additionally, the correlation between experimental and predicted diameters in both training and testing data of the ANN model exhibited high reliability. The error values for training and testing datasets were closely aligned (Table 2). The Total Generalization Value (TGV) of 1.87 in training data, along with a value close to 2 in testing data, confirms the suitability of this model.

One of the key strengths of the ANN model is the similarity between training and testing data, which underscores its robustness. Given its ability to handle non-linear relationships, the ANN model proves to be more suitable for predicting the optimal nanofiber diameter compared to the RSM model. Moreover, Table 6S presents the weights and bias values of the neural network. (Eq. (14)) was utilized to assess the impact of each parameter on the nanofiber diameter (output variable). The parameter with the highest value (input variable) had the greatest influence on nanofiber diameter. Fig. 8 illustrates the percentage effect of different parameters on nanofiber diameter in the ANN model. The results indicate that PXDDA polymer concentration (77.9 %) had the most significant impact, followed by voltage (16.9 %), which played a greater role than flow rate. These relationships are consistent with the RSM model.

In the Genetic Algorithm (GA) model, the cost values of 20 samples were calculated (Table 3). The results indicate that samples with lower cost values were closest to the target data. According to Table 3, Sample 2 had a cost value of 0.0054 and a nanofiber diameter of 271 ± 70 nm, making it the most suitable sample, closely matching the target data. Furthermore, the GA model predicted the optimal nanofiber diameter of 269 nm, fabricated using 16% (w/w) PXDDA polymer solution, 14 kV applied voltage and 0.8 mL/h flow rate. This prediction closely aligns with the desired nanofiber diameter. The ANN model, utilizing non-linear relationships, provided a more accurate prediction of nanofiber diameter compared to the RSM model, making it a better approach for optimization.

**Table 2.** The performance of the ANN models during the training and testing steps to analysis of PXDDA/PCL/Gelatin nanofiber diameter.

| Parameter | Diameter | |
|---|---|---|
| | **Train** | **Test** |
| MSE | 0.00033 | 0.0131 |
| $R^2$ | 0.9939 | 0.8168 |
| GV | 1.9329 | 1.6247 |
| TGV | 1.8713 | 1.8713 |

**Table 3.** The cost value of samples with changing parameters to fabricate PXDDA/Gel/PCL nanofibers.

| Input variable | | | | Output variable | |
|---|---|---|---|---|---|
| Run | $X_1$ Concentration (%) | $X_2$ Voltage (kV) | $X_3$ Flow rate (ml/h) | Diameter (nm) | Cost value |
| 1 | 11.5 | 20 | 0.7 | 211 ± 36 | 0.2178 |
| 2 | 15 | 15 | 0.5 | 271 ± 70 | 0.0054 |
| 3 | 20 | 20 | 0.7 | 351 ± 95 | 0.3063 |
| 4 | 25 | 15 | 0.5 | 367 ± 112 | 0.3666 |
| 5 | 28.4 | 20 | 0.7 | 478 ± 192 | 0.7822 |
| 6 | 15 | 25 | 1 | 257 ± 51 | 0.0452 |
| 7 | 20 | 20 | 1.1 | 357 ± 83 | 0.3297 |
| 8 | 20 | 12.9 | 0.7 | 289 ± 49 | 0.0718 |
| 9 | 20 | 20 | 0.7 | 360 ± 97 | 0.3381 |
| 10 | 20 | 20 | 0.7 | 354 ± 95 | 0.3171 |
| 11 | 20 | 20 | 0.7 | 357 ± 82 | 0.3279 |
| 12 | 20 | 20 | 0.3 | 354 ± 107 | 0.3153 |
| 13 | 20 | 27 | 0.7 | 334 ± 71 | 0.2401 |
| 14 | 15 | 25 | 0.5 | 227 ± 60 | 0.1584 |
| 15 | 25 | 25 | 1 | 430 ± 105 | 0.6005 |
| 16 | 20 | 20 | 0.7 | 352 ± 91 | 0.31 |
| 17 | 15 | 15 | 1 | 230 ± 53 | 0.1465 |
| 18 | 25 | 15 | 1 | 370 ± 86 | 0.3774 |
| 19 | 25 | 25 | 0.5 | 434 ± 67 | 0.6158 |
| 20 | 20 | 20 | 0.7 | 350 ± 96 | 0.3008 |

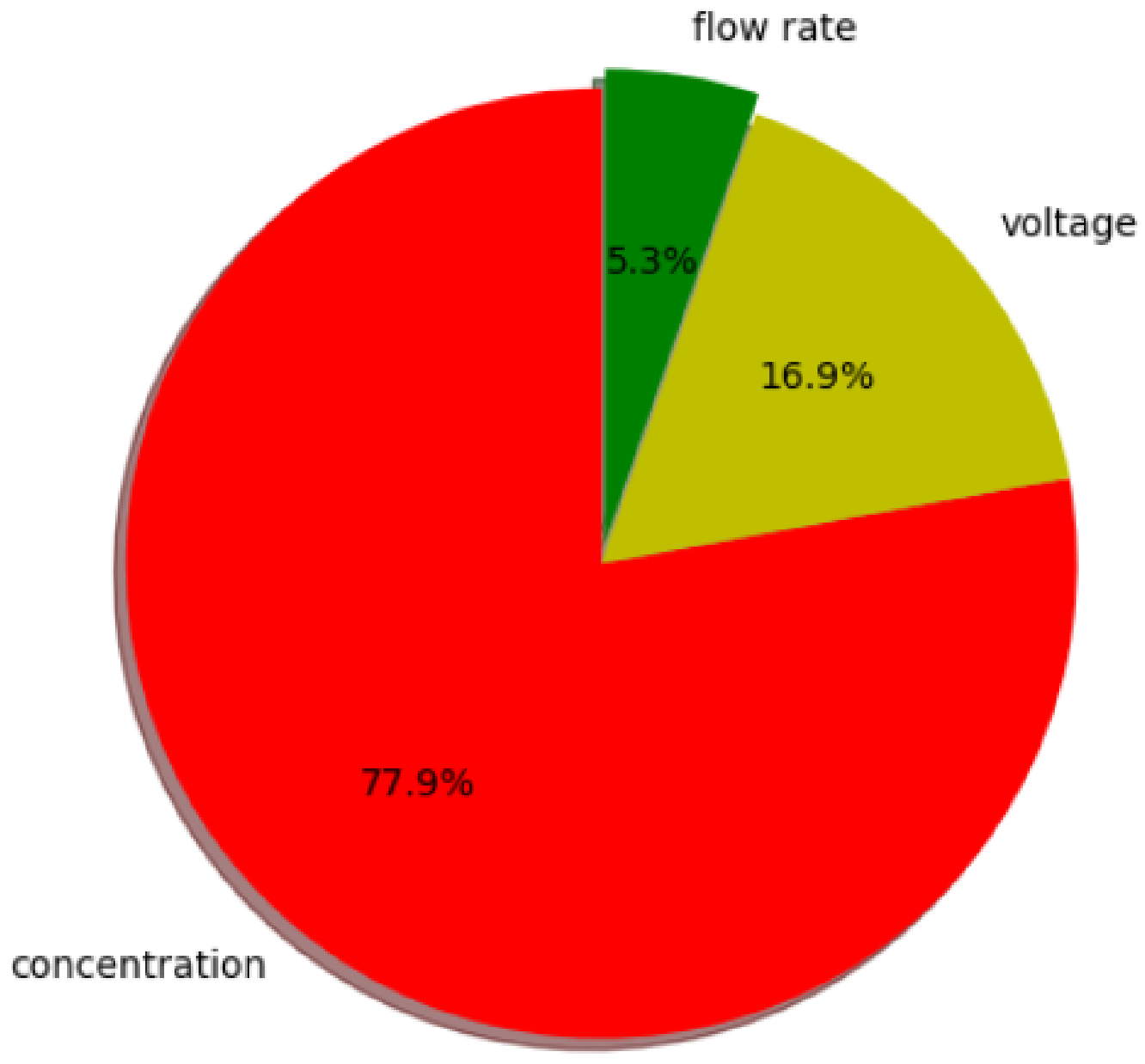

**Fig. 8.** The percentage of effective different parameters on the diameter of nanofibers in ANN model.

$$R_i(\%) = \sum_{j=1}^{n_H}\left(\left|W_{ij}^{(1)}\right|\left|W_j^{(2)}\right| / \sum_{l=1}^{n_I}\left|W_{lj}^{(1)}\right|\right) / \sum_{i=1}^{n_I}\left(\sum_{j=1}^{n_H}\left(\left|W_{ij}^{(1)}\right|\left|W_j^{(2)}\right| / \sum_{l=1}^{n_I}\left|W_{lj}^{(1)}\right|\right)\right) \times 100 \quad (14)$$

### 3.3. Physical and biological of bilayer nanofibers

#### 3.3.1. Morphology of one layer crosslinked Nanofibers

To improve their stability in water, PXDDA/Gel/PCL nanofibrous membranes were chemically crosslinked. One method involved a glutaraldehyde (GA) solution as a crosslinking agent. The primary goal was to crosslink PXDDA/Gel nanofibers, which in turn facilitated crosslinking of PXDDA/Gel/PCL nanofibers. Therefore, examining crosslinking durations for PXDDA/Gel mats helped determine the optimal time for PXDDA/Gel/PCL nanofiber crosslinking. Fig. 9 presents SEM images showing nanofiber morphology following crosslinking with GA/HCl vapor treatment at different times. The results demonstrated that PXDDA/Gel nanofibers did not crosslink at 0.5 or 1 hour (Fig. 9(a-b)). The optimal crosslinking time for PXDDA/Gel nanofibers was determined to be 2.5 hours (Fig. 9(c)), while PXDDA/Gel/PCL mats started crosslinking at 2 hours (Fig. 9(d)). The optimal crosslinking time for PXDDA/Gel/PCL mats without structural degradation was 2.5 hours in the oven (Fig. 9(e)). However, extending the crosslinking time beyond 2.5 hours resulted in the weakening of the chemical structure and reduction in nanofiber diameter, as shown in Fig.

9(f-g). Additionally, the morphology of crosslinked PXDDA/Gel/PCL nanofibers was examined following the spin coating of CD/N-ACNP capsules as antibacterial agents, presented in Figure 10.

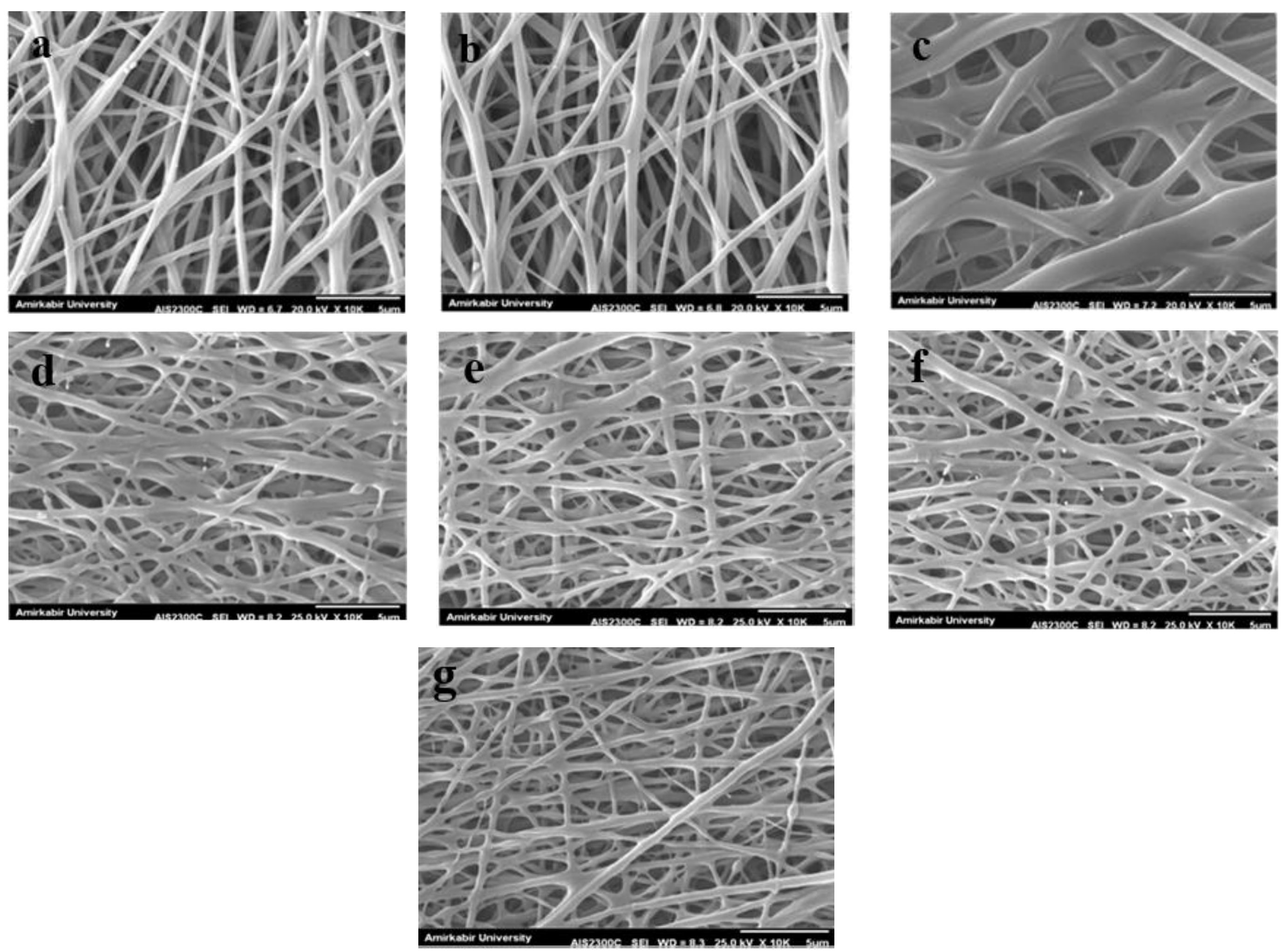

**Fig. 9.** Morphology of PXDDA/Gel nanofibers in the oven for different times a) 0.5 h, b) 1 h, c) 2 h, and morphology of PXDDA/Gel/PCL nanofibers in the oven for different times d) 2 h, e) 2.5 h, f) 3 h, and g) 4 h.

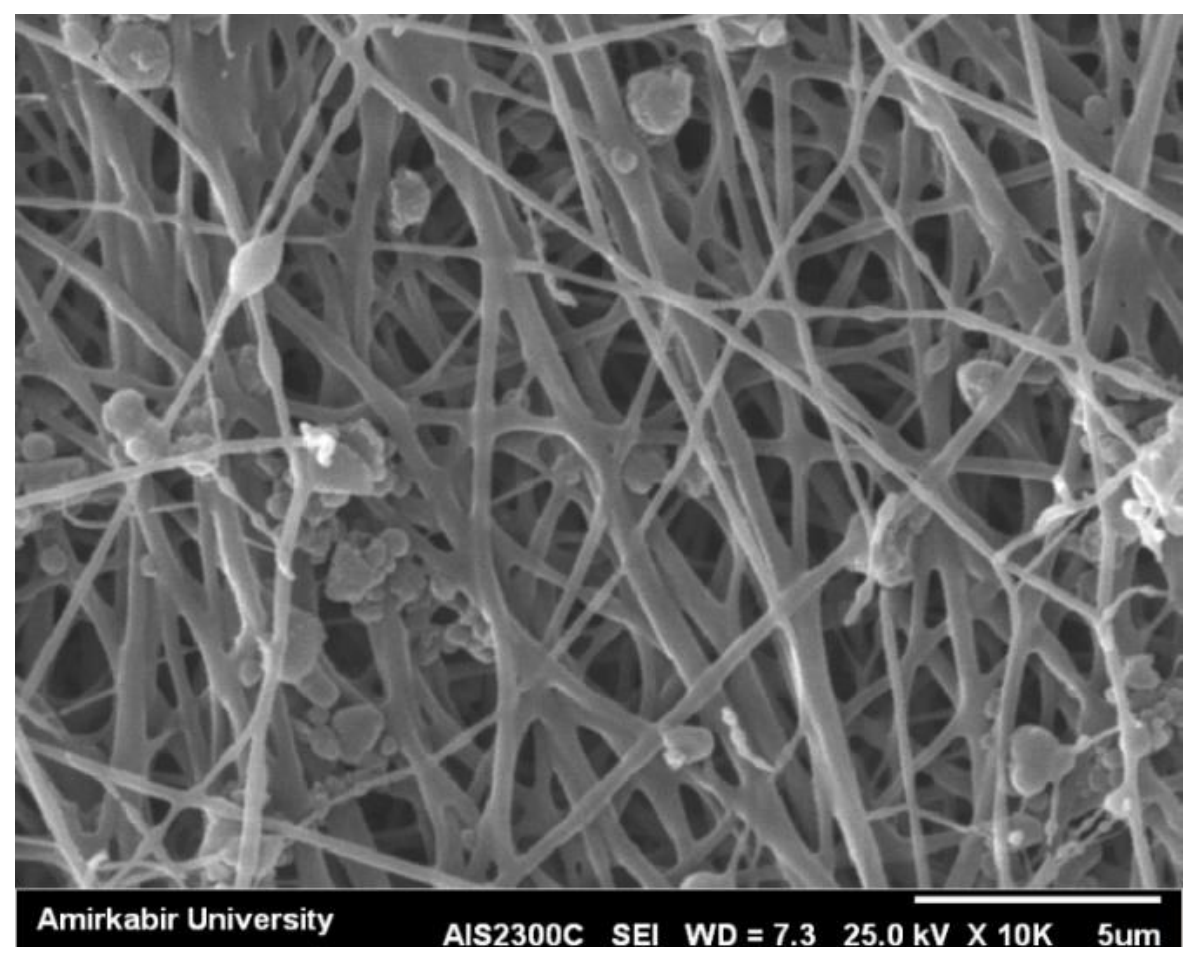

**Fig. 10.** Morphology of capsule on the PXDDA/Gel/PCL crosslinked nanofibers.

### 3.3.2. FTIR Analysis

FTIR technique was used to confirm the functional groups in different biomaterials which are shown in Fig. 11. The peaks of the CD include 915 $cm^{-1}$ (C-Cl groups), 1023 $cm^{-1}$ (C – C), 1149 $cm^{-1}$ (C – O – C), 1378 $cm^{-1}$ (SC-H), and 1459 $cm^{-1}$ (C-N groups), 2943 $cm^{-1}$ (CH), and 3311 $cm^{-1}$ (OH, NH) that were mixed with N-ACNP. Heris et al. evaluated the presence of peaks at 1532 $cm^{-1}$, 1466 $cm^{-1}$, 1371 $cm^{-1}$, and 818 $cm^{-1}$ in the CD that indicated amide-type C=O, C-N, SC-H, and C-Cl groups, respectively [37]. Also, N-ACNP has peaks in wave numbers of 1124 $cm^{-1}$, 1574 $cm^{-1}$, 2046 $cm^{-1}$, 2921 $cm^{-1}$, and 3438 $cm^{-1}$ including C-O-C groups, amide III C-N stretching vibrations, C≡ N groups, C ≡ C group, and N-H groups, respectively. The incorporation of CD into N-ACNP enhances bond absorptions in their wave numbers which are shown in Fig. 11(a).

The results demonstrate the different groups of PXDDA/Gel nanofibers that were incorporated together. The peak 3327 $cm^{-1}$ belongs to N-H stretching of amides I and II in gelatin, and 2944 $cm^{1}$ is assigned to symmetrical and asymmetric C-H stretching. Peaks of 1732 $cm^{-1}$, and 1652 $cm^{-1}$ include C=O stretching in PXDDA polymer, and N-H bending of amide I in gelatin structure, respectively. Also, peaks of 1542 $cm^{-1}$, and 1456 $cm^{-1}$ confirm functional groups including combination N-H bending, C-N stretching of amide II, and combination COO group and aliphatic groups, in gelatin polymer, respectively. Moreover, asymmetric and symmetric C – O – C stretching have been exhibited at peaks of 1242 $cm^{-1}$, and 1056 $cm^{-1}$ that are present in PXDDA polymer. Furthermore, PCL polymer has different functional groups that include 2946 $cm^{-1}$ (symmetrical and asymmetric CH stretching), 1731 $cm^{-1}$ (C=O stretching), 1240 $cm^{-1}$, and 1175

cm$^{-1}$ (asymmetric and symmetric C – O – C stretching). Sharhai et al. confirmed functional groups of gelatin and PCL polymer in similar peaks [33]. The incorporation of PXDDA, Gel, and PCL polymer solutions to fabricate nanofibers caused the creation of peaks of absorption spectrums similar to other peaks that are described. The peak of GA for crosslinking of nanofibers at 1583 cm$^{-1}$ (C=N group) was merged with 1542 cm$^{-1}$ of gelatin polymer which is shown in Fig. 11(b). Finally, all of the chemical structures in these materials are incorporated together and their peaks have been described. Some of these peaks were shifted to a lower wavelength and some of them appeared in higher wavelength.

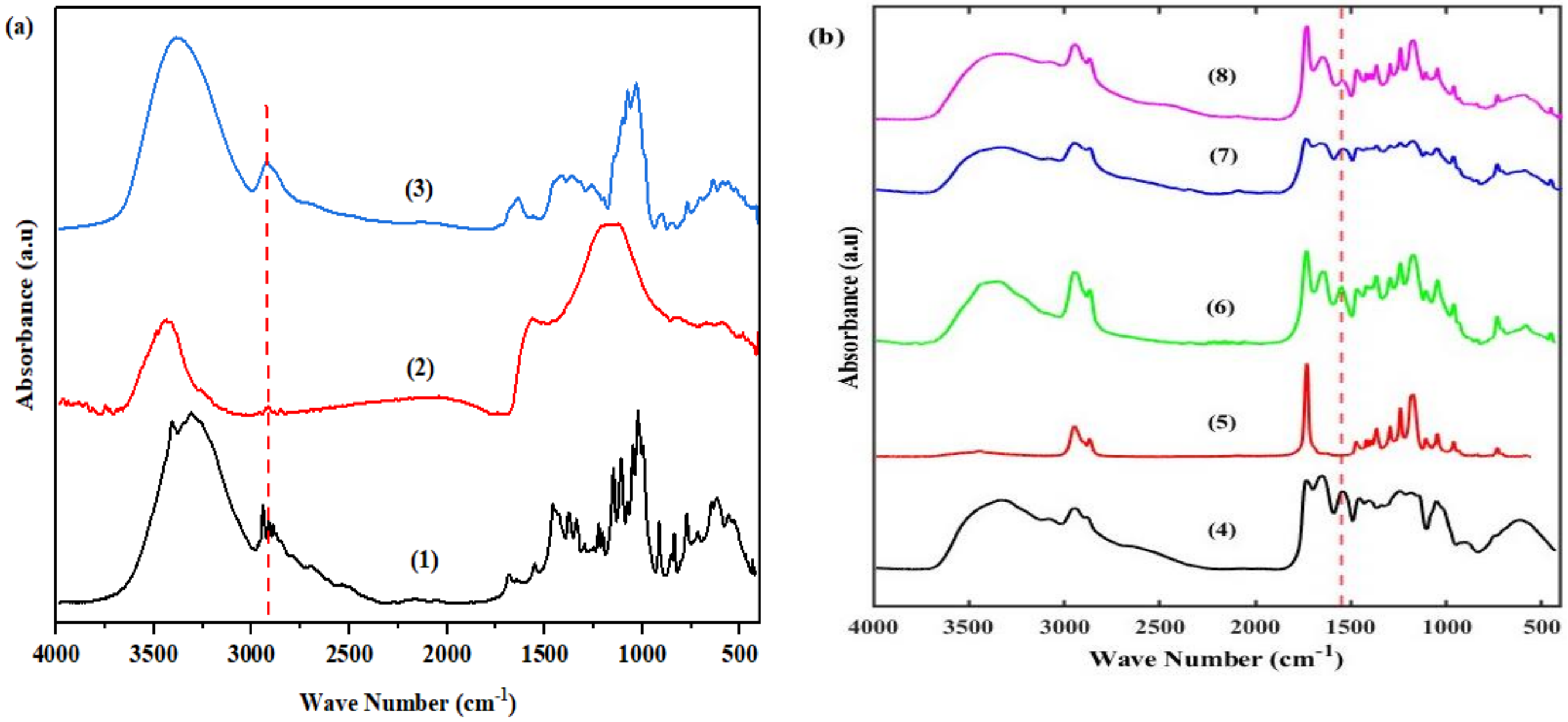

**Fig. 11.** FTIR images of a) powders including 1) Clindamycin drug, 2) N-ACNP, 3) Capsule, b) nanofibers including 4) PXDDA/Gel, 5) PCL, 6) One layer of PXDDA/Gel/PCL without crosslinked nanofibers, 7) One layer of PXDDA/Gel/PCL crosslinked nanofibers, 8) Bilayer of PXDDA/Gel/PCL crosslinked nanofibers.

### 3.3.3. X-ray Diffraction (XRD) Analysis

X-ray diffraction patterns were done on N-ACNP, capsule, one layer of nanofibers without crosslinking. Also, the effect of CD concentration in capsules between layers of nanofibers on microstructure that are shown in Fig. 10S. N-ACNP has a broad, high- intensity peak around 20-25°, and another peak at 44° with crystallinity about 49.66%. Also, the capsule indicates a moderate decrease in structural order and yielding a crystallinity of 45.77%. Though CD has crystallinity structure, the capsule had lower crystallinity than CD, due to molecular dispersion in the N-ACNP. Additionally, Wu et al. analyzed XRD patterns of various carbon samples. Their findings revealed that carbon particles (CPs) exhibit a broad, high-intensity peak around 22°, indicating amorphous carbon. Furthermore, incorporating urea into CP structures did not alter their configuration. However, heat activation shifted the XRD peak to 24°, while a new peak at 44° formed, corresponding to (002) and (100) lattice spacings, indicating a graphitic phase [30]. Conversely, Ranajikiri et al. examined the crystalline state of CD, identifying intense peaks at 5.0°, 8.7°, 10.1°, 11.5°, 13.6°, 15.4°, 17.8°, and 23.1° [38].

For the one layer of nanofibers, a sharper diffraction peak near 22.5° was observed, associated with the semi-crystalline nature of polycaprolactone (PCL). The crystallinity index increased to 48.73%**,** indicating partial phase separation and local ordering in the polymer matrix. In contrast, the bilayer nanofibers with CD-coated capsules displayed relatively broad diffraction peaks around 20-25°. The presence of a capsule between nanofiber layers resulted in the absence of the crystalline nature of CD. Within the nanofiber structure, PXDDA is a synthetic polymer with an amorphous structure, Gelatin is a natural polymer with an amorphous structure, lacking distinct peaks in the XRD pattern. However, the absence of sharp, well-resolved peaks indicates that the structures remain semi-crystalline rather than fully crystalline**.**

#### 3.3.4. Surface Hydrophilicity of Nanofibers

Water absorption on the surface of wound dressing mats plays a crucial role in absorbing fluids, including exudates from injured and inflamed tissues. Additionally, nanofiber hydrophilicity is essential for uniform cell distribution within the scaffold and for ensuring efficient oxygen and nutrient exchange between the scaffold and its environment [39]. Nanofiber hydrophilicity was analyzed by measuring water absorption on the nanofiber surface. Due to the hydrophilic nature of PXDDA/Gel nanofibers, their contact angle was 0°. To reduce hydrophilicity, PCL polymer was incorporated into PXDDA/Gel nanofibers, increasing the contact angle to 44 ± 7.1° (Fig. 12(a)). To stabilize water absorption, glutaraldehyde (GA) vapor was used for crosslinking nanofiber surfaces, which increased the contact angle to 55 ± 2.5° due to reactions between hydrophilic groups (amino groups) and the crosslinking agent (aldehyde groups) (Fig. 12(b)). Following nanofiber surface modification with spin-coated capsules, a second nanofiber layer was electrospun. GA vapor was then applied to crosslink the second layer, regulating drug release.

Initially, contact angle measurements were conducted on the first layer of crosslinked nanofibers, coated with capsules containing different CD concentrations (Fig. 12(c-e)). Capsule coating on nanofiber surfaces reduced hydrophilicity, primarily due to the conjugation of hydrophilic groups between CD and N-ACNP and CD entrapment within N-ACNP. Additionally, high-temperature activation of N-ACNP led to hydrogen atom reduction, forming N-H and C≡N groups and increasing aromatic ring content. This resulted in an increased contact angle of approximately 71 ± 0.7°. Conversely, increasing CD content led to a decrease in contact angle due to the presence of free CD on the nanofiber surface, forming hydrogen bonds between hydroxyl groups in CD and water molecules.

Hydrophilicity was also evaluated in bilayer nanofibers, where polymer solutions were electrospun onto capsules and subsequently crosslinked using GA vapor (Fig. 12(f-h)). The results indicated that nanofiber hydrophilicity increased at lower CD concentrations in N-ACNP, compared to single-layer nanofibers, because most CD was entrapped within N-ACNP, triggering water absorption and burst drug release. Consequently, the contact angle decreased, as shown in Table 7S. However, higher CD concentrations resulted in an increased contact angle, likely due to enhanced drug entrapment within N-ACNP and formation of emulsions through hydrophilic group interactions, which led to burst drug release and further contact angle increase. Setia Budi et al. examined the hydrophilicity of PCL/Gel nanofibers loaded with clindamycin for tissue engineering applications. Their findings revealed that incorporating graphene oxide (GO) and clindamycin into PCL/Gel nanofibers decreased the contact angle from 87.6° to 49.6° [46].

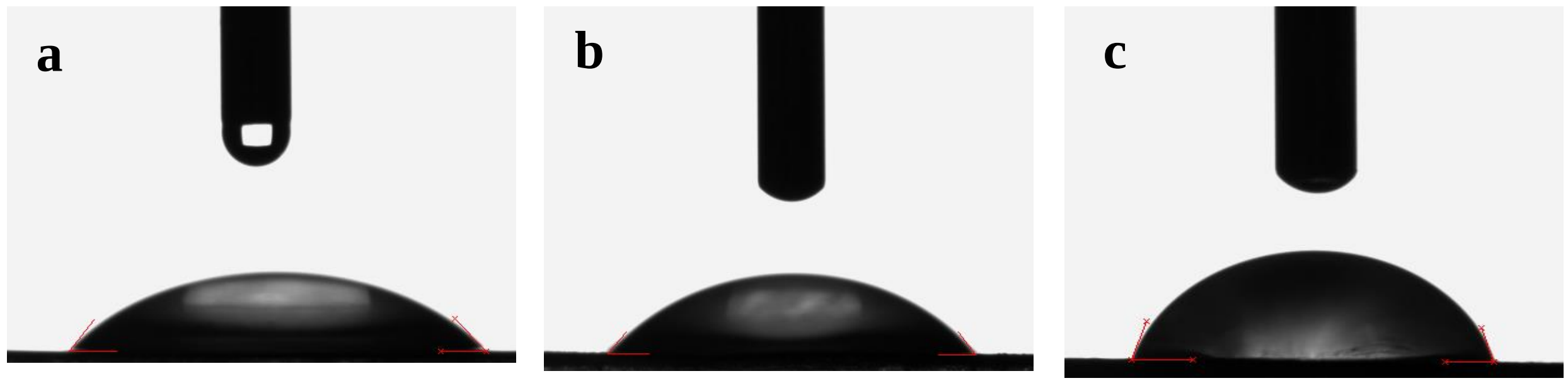

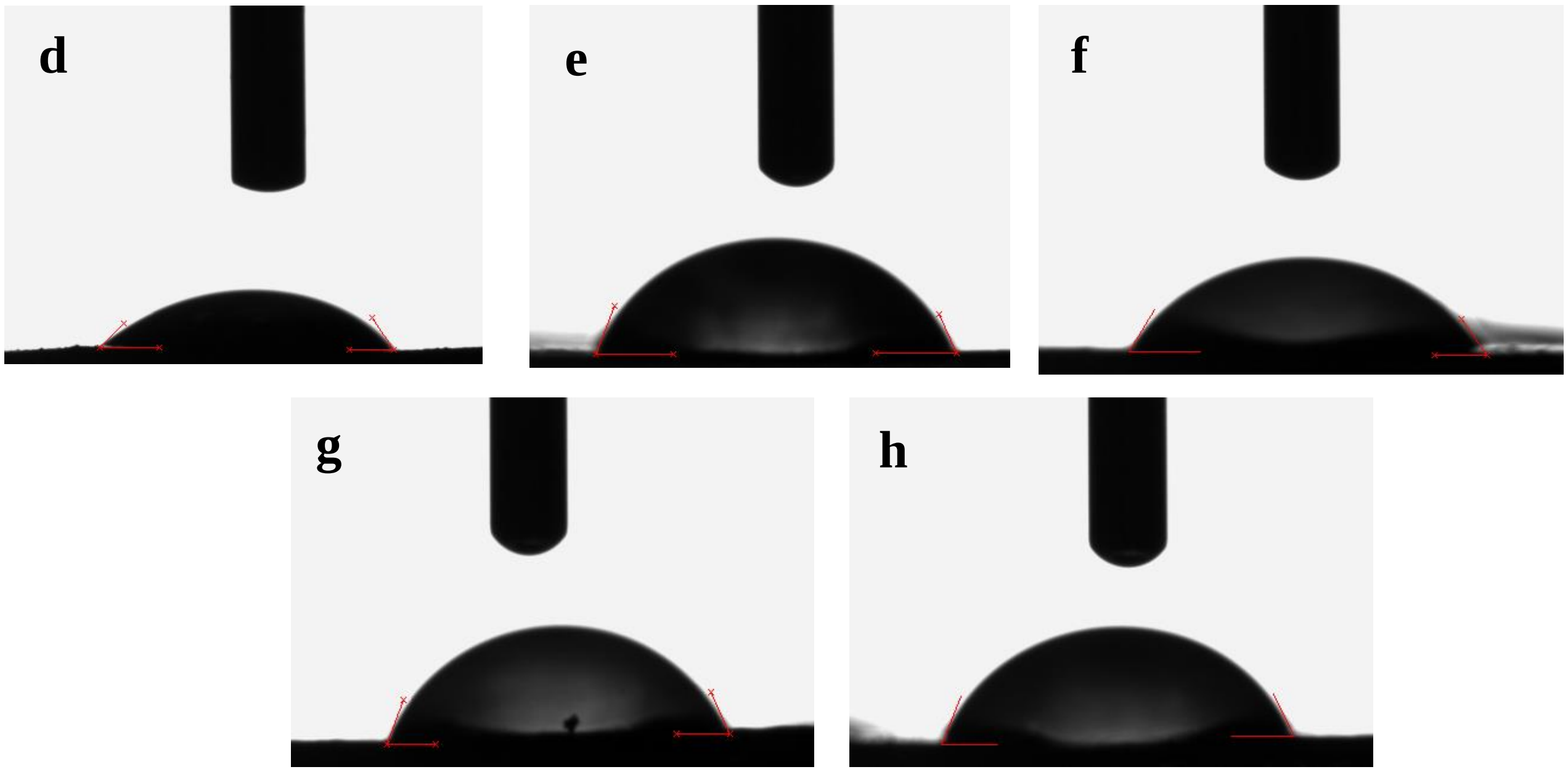

**Fig. 12.** Contact angle of nanofibers a) PXDDA/Gel/PCL nanofibers, b) PXDDA/Gel/PCL crosslinked nanofibers for 2.5 hours, One layer of PXDDA/Gel/PCL nanofibers with coating CD/N-ACNP with a concentration of c) (0.0005/0.001) g/mL d) (0.001/0.001) g/mL, e) (0.002/0.001) g/mL on them, Bilayer of PXDDA/Gel/PCL nanofibers with coating CD/N-ACNP with a concentration of f) (0.0005/0.001) g/mL, g) (0.001/0.001) g/mL, and h) (0.002/0.001) g/mL on them.

### 3.3.5. Bulk Hydrophilicity of Nanofibers

The absorption of moisture in wound dressings is critical for enhancing their properties and effectiveness in the wound-healing process. To evaluate the effect of capsules on water absorption capacity, the swelling behavior of the samples was studied. Fig. 11S illustrates the swelling percentage of nanofibers after immersion in phosphate-buffered saline (PBS) for 12 hours. The results indicate that loading capsules between nanofiber layers led to increased water absorption over different time points before sample degradation began. Specifically during the first 12 hours, the addition of CD (at all concentrations) into N-ACNP enhanced the swelling percentage of nanofibers. However, prolonged immersion (beyond 12 hours) resulted in sample degradation, triggering a reduction in weight and swelling percentage, particularly after 24 hours. Increasing CD concentration within N-ACNP further raised the contact angle while reducing the swelling percentage, compared to lower CD concentrations. This phenomenon is likely due to CD emulsion formation on the nanofiber surface, which limited the permeability of water molecules through scaffold pores.

ANOVA analysis showed no significant difference in swelling percentage between samples with different capsule concentrations across time points. The CD content did not have a considerable impact on the swelling percentage overall. However, the swelling percentage between 0.5 hours and 12 hours was found to be statistically significant ($p < 0.001$), confirming its importance in water absorption compared to other time intervals.

### 3.3.6. In Vitro Clindamycin Drug Release

The release of clindamycin (CD) drug was assessed from a bilayer of PXDDA/Gel/PCL nanofibers, demonstrating its ability to control drug release from scaffolds. Additionally, loading CD into N-ACNP helped prevent burst drug release, ensuring a sustained release for effective wound healing over suitable time intervals. Coating CD without a capsule between nanofiber layers resulted in a sudden and rapid drug release (~60%) within the first three hours. Fig. 13 presents the cumulative release profile of CD-loaded N-ACNP in varying concentrations (0.0005, 0.001, and 0.002 g/mL) within a bilayer of crosslinked nanofibers over 72 hours.

The results illustrate that all release curves followed a similar trend. The initial burst release behavior suggests weak interactions between drug molecules and N-ACNP, or poor drug diffusion into particles as drug concentration increased. Higher drug concentrations exceeded N-ACNP capacity, limiting drug diffusion and triggering increased initial burst release. Conversely, the initial burst release rate was slower across all drug concentrations, attributed to the bilayer structure of crosslinked nanofibers. The maximum cumulative drug release recorded was 62% release at 0.0005 mg/mL CD, 73% release at 0.001 mg/mL CD, and 90% release at 0.002 mg/mL CD.

Encapsulation of CD into N-ACNP pores helped regulate drug release from nanofibers. Immersing the samples in PBS led to nanofiber swelling in the release media, reaching maximum water uptake within 12 hours. In the subsequent stage N-ACNP and PXDDA/Gel/PCL nanofiber degradation facilitated drug release into the scaffold. The release process transitioned into sustained release, primarily driven by CD loading into N-ACNP, nanofiber matrices, and crosslinking agents.

In the final days, the drug release rate slowed, resulting in low-gradient drug diffusion. Also, similar results were obtained from other studies about drug release rate in multilayer nanofibers and loading drugs into nanoparticles that attributed to sustaining drug delivery [36,40].

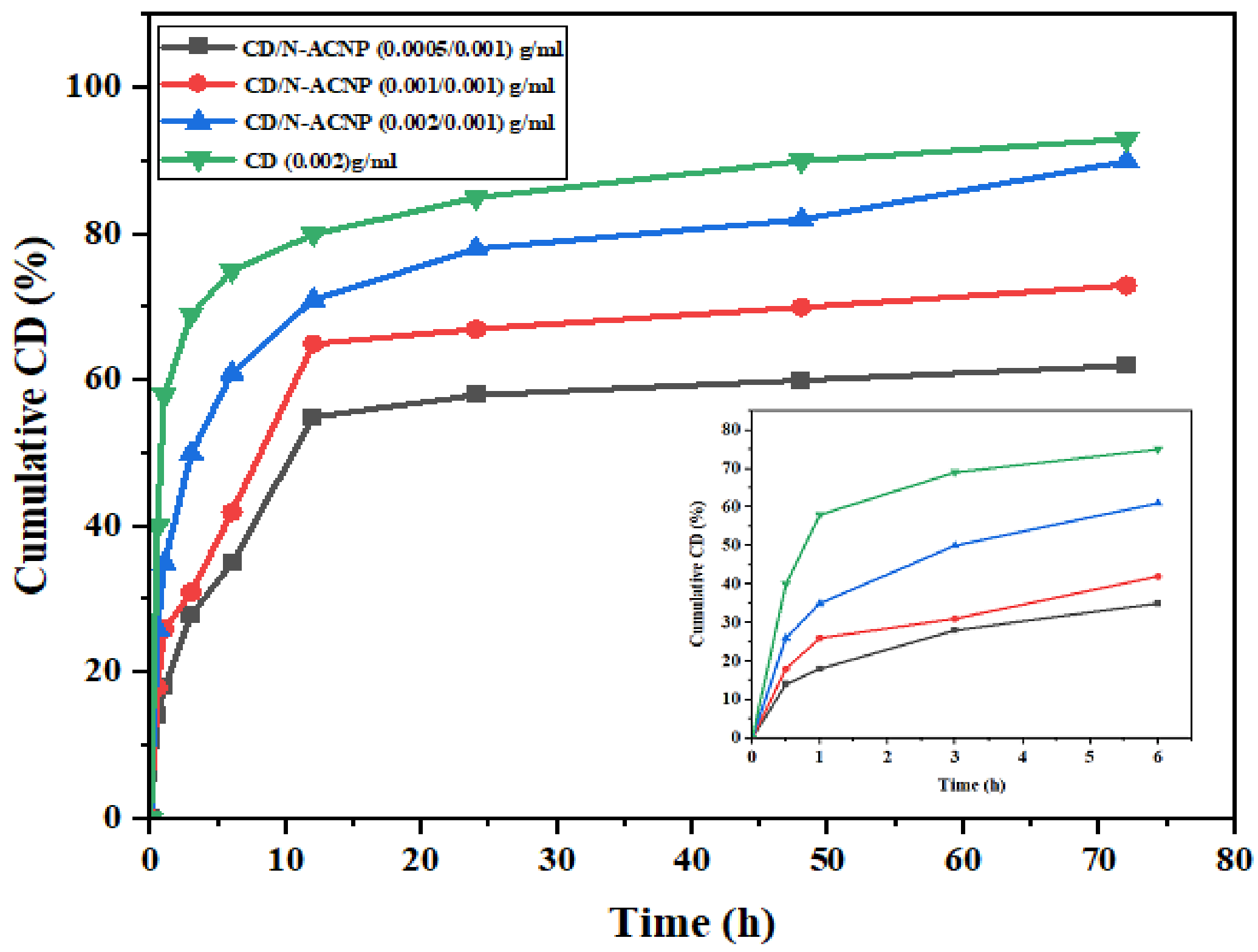

**Fig. 13.** The release profile of CD from a bilayer of PXDDA/Gel/PCL crosslinked nanofibers.

### 3.3.7. Mechanisms of Drug Release

Several models were employed to investigate the clindamycin (CD) release mechanism. The Korsmeyer-Peppas kinetic model was applied to analyze the diffusion mechanism. In this model, the release exponent (n) determines the drug release mechanism. If $n < 0.45$ is related to Fickian diffusion, $0.45 < n < 1$ is related to Non-Fickian diffusion, $n = 1$ is related to Zero-order kinetics or Case-II transport, and $n > 1$ is related to Super Case-II transport [16]. Table 4 presents the calculated release constants, release exponent (n), and regression coefficients ($R^2$). The results indicate that all samples exhibited high correlation values, adhering to the Higuchi kinetics model as the best-fit method. Furthermore, Korsmeyer-Peppas equation analysis revealed sample 1 (0.0005 g/mL CD concentration) exhibited $0.45 < n < 1$, indicating a non-Fickian diffusion mechanism. Other samples with $n < 0.45$ followed a Fickian diffusion mechanism. These drug release mechanisms are influenced by various parameters, including crosslinking factor, CD concentration, swelling degree, and diffusion distance.

**Table 4.** Kinetic of CD release from a bilayer of nanofibers in four mathematical models.

| Samples | Zero-order kinetic model | | First-order kinetic model | | Higuchi model | | Korsemeyer-Peppas model | | |
|---|---|---|---|---|---|---|---|---|---|
| | $K_0$ | $R^2$ | $K_t$ | $R^2$ | $K_{HC}$ | $R^2$ | n | $K_{KP}$ | $R^2$ |
| 1 | 1.227 | 0.9052 | 0.0154 | 0.5863 | 10.709 | 0.958 | 0.6318 | 2.0964 | 0.8679 |
| 2 | 0.7458 | 0.3808 | 0.0065 | 0.2991 | 8.7319 | 0.6071 | 0.3605 | 2.7446 | 0.8044 |
| 3 | 1.1118 | 0.901 | 0.0105 | 0.7894 | 10.646 | 0.961 | 0.3866 | 2.5183 | 0.9469 |
| 4 | 0.3144 | 0.2631 | 0.0021 | 0.219 | 3.7401 | 0.4331 | 0.1217 | 3.05 | 0.617 |

**1: CD/N-ACNP (0.0005/0.001) g/mL**

**2: CD/N-ACNP (0.001/0.001) g/mL**

**3: CD/N-ACNP (0.002/0.001) g/mL**

**4: CD (0.002) g/mL**

### 3.3.8. Biodegradability of Nanofibers

The degradation behavior of polyester biopolymers is a key factor in assessing their suitability for tissue engineering. Swelling and degradation properties play a critical role in wound healing, as they facilitate oxygen and nutrient diffusion and promote cell adhesion [41,42]. Fig. 12S illustrates a hydrolytic degradation test conducted on nanofiber samples at pH = 7, examining the effects of PBS absorption on nanofiber morphology, analyzed using SEM imaging. To control and slow degradation, PCL polymer was incorporated into PXDDA/Gel polymer solutions, and nanofibers were fabricated via electrospinning. Additionally, bilayer wound dressings were crosslinked with GA, improving mechanical stability within scaffolds. Despite this approach, PXDDA and gelatin polymers exhibited rapid degradation. PCL incorporation in an equal molar ratio had little effect on reducing degradation, due to higher hydrophilic content compared to hydrophobic groups in the polymer. Water molecules rapidly hydrolyzed ester bonds, diffusing into scaffolds. The control sample (a single-layer PXDDA/Gel/PCL crosslinked nanofiber) completely degraded within 10 days. Bilayer crosslinked nanofibers with capsules had a slower degradation rate than the control sample, due to loading capsules between layers of nanofibers according to section 3.4, and section 3.5. Water absorption and swelling in nanofiber structures contributed to crosslink degradation, ultimately releasing capsules from bilayer nanofibers. Increasing drug loading into capsules resulted in higher swelling rates and structural degradation over 30 days.

### 3.3.9. Antibacterial Properties

The antibacterial properties of N-ACNP and bilayer PXDDA/Gel/PCL crosslinked nanofibers coated with capsules in various concentrations between layers of them were evaluated against Staphylococcus aureus (S. aureus) and Escherichia coli (E. coli) using disk diffusion tests and bacterial colony suspension assays. To assess bacterial inhibition, 100 µL of bacterial suspension ($1.5 \times 10^8$ CFU/mL) was spread onto an agar plate. The results showed that N-ACNP, similar to bacterial suspensions, exhibited no inherent antibacterial properties despite high-temperature activation for nanoparticle synthesis and urea-induced amide bonding formation. Bilayer PXDDA/Gel/PCL nanofibers coated with capsules between layers of them displayed antibacterial activity. The highest drug concentrations within nanofiber layers resulted in inhibition zone diameters of 20 mm (S. aureus) and 22 mm (E. coli). The effect of E. coli inhibition was more pronounced at higher drug concentrations, compared to S. aureus are shown in Table 8S.

Moreover, samples with varying capsule concentrations were immersed in 1 mL of bacterial suspension ($1.5 \times 10^8$ CFU/mL) for 2 hours. The results showed that increasing drug concentrations (0.0005 g/mL to 0.002 g/mL) resulted in the reduction of bacterial count from $4.2 \times 10^5$ to 0 against E. coli and a reduction of bacterial count from $6.1 \times 10^6$ to 0 against S. aureus. According to bacterial colony suspension assays, E. coli bacterial growth was lower than S. aureus bacterial growth (Table 5).

**Table 5.** Antibacterial properties of N-ACNP, and effective capsule in different concentrations of CD in the bilayer of nanofibers.

| Sample | (E.coli) | (S.aureus) | ( Cuf /mL) | |
|---|---|---|---|---|
| | | | **E.coli** | **S.aureus** |
| **Control CD 0 (g/mL)** | | | **$1.5\times10^{8}$** | **$1.5\times10^{8}$** |
| **N-ACNP** | | | **$1.5\times10^{8}$** | **$1.5\times10^{8}$** |
| **CD 0.0005 (g/mL)** | | | **$4.2\times10^{5}$** | **$6.1\times10^{6}$** |

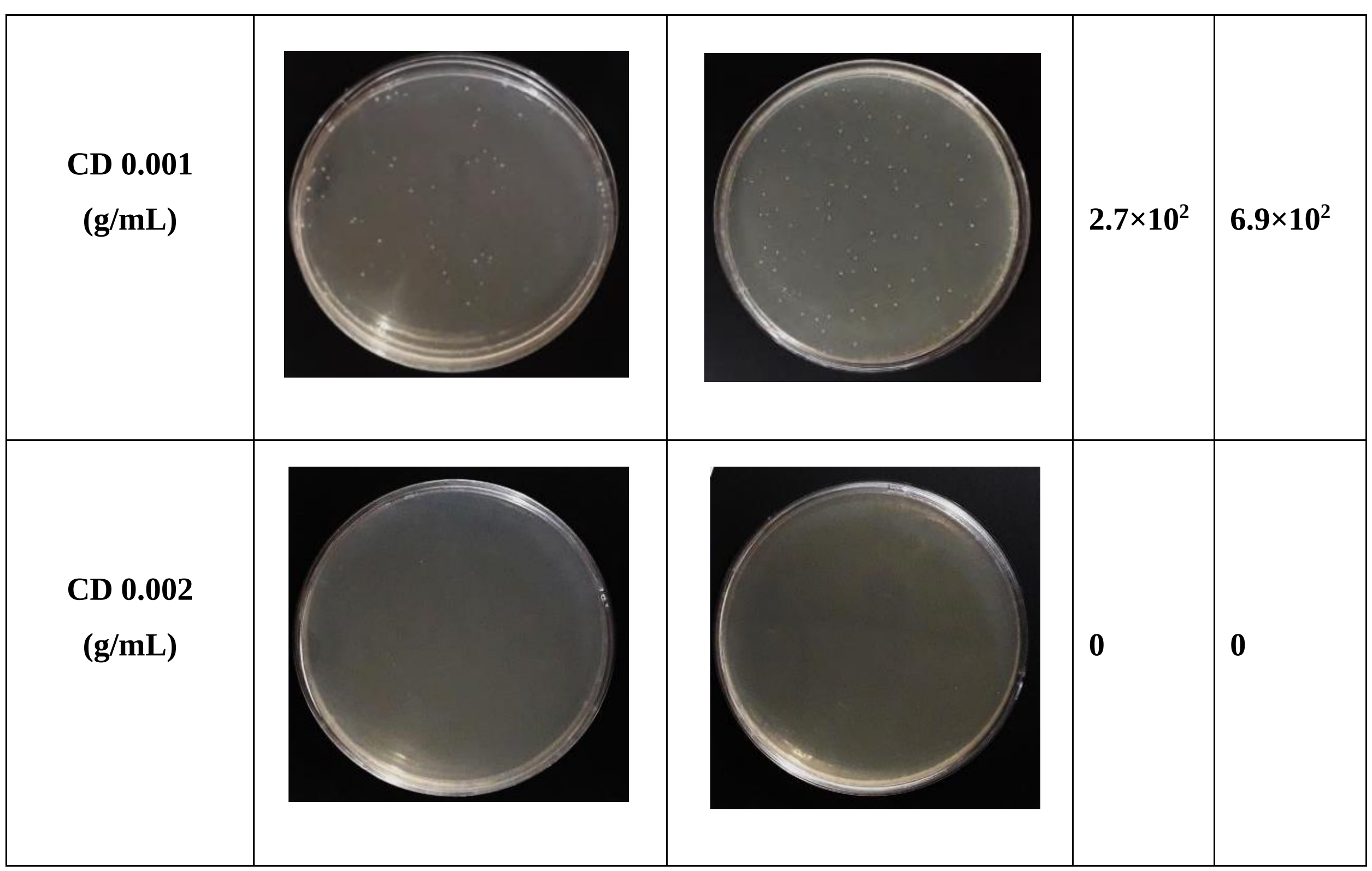

| | | | | |
|---|---|---|---|---|
| **CD 0.001 (g/mL)** | | | **$2.7\times10^2$** | **$6.9\times10^2$** |
| **CD 0.002 (g/mL)** | | | **0** | **0** |

### 3.3.10. Cell Culture and MTT Assay

The cytotoxicity of nanofibers was assessed using mesenchymal stem cells (MSCs) via the MTT assay. Analysis of cell viability data demonstrated that capsule-coated nanofibers exhibited slight toxicity over three days. For the control sample, the metabolic activity of untreated cells was considered 100%. Comparatively, wound dressings incorporating capsules in different drug concentrations exhibited significantly higher cell viability. On day 3, the sample containing 0.001 g/mL of CD exhibited the highest cell viability (~95%), as shown in Fig. 14 (A). However, increasing CD concentration between nanofiber layers led to greater toxicity, attributed to antibiotic drug release. Therefore, determining the optimal CD concentration is crucial for maintaining cell viability over three days. Despite this, all capsule-loaded samples showed higher cell viability than the control.

The most significant difference was observed between 0.002 g/mL CD in bilayer nanofibers and the control sample, with $p < 0.0001$ and $p < 0.001$ across all days. No statistical differences in cell viability were observed between days. Thus, CD concentration was the most influential factor affecting cell viability, rather than time. Karczewski et al. examined cell viability following the loading of clindamycin and triple antibiotic into polydioxanone nanofibers. Their study found that higher drug concentrations led to increased toxicity, rising from 52% on day 1 to 63% on day 28 [43].

The morphology and spreading of MSCs were examined after 24 hours using SEM, as shown in Fig. 14(B). The presence of biocompatible polymers (PCL, gelatin, and PXDDA) within the nanofiber structure facilitated cell attachment on PXDDA/Gel/PCL nanofibers (control sample), which lacked capsules. However, capsule-loaded bilayer nanofibers exhibited enhanced cell adhesion compared to the control sample. Specifically, the 0.001 g/mL CD concentration in N-ACNP demonstrated optimal cell spreading, correlating with higher cell viability and improved cell growth.

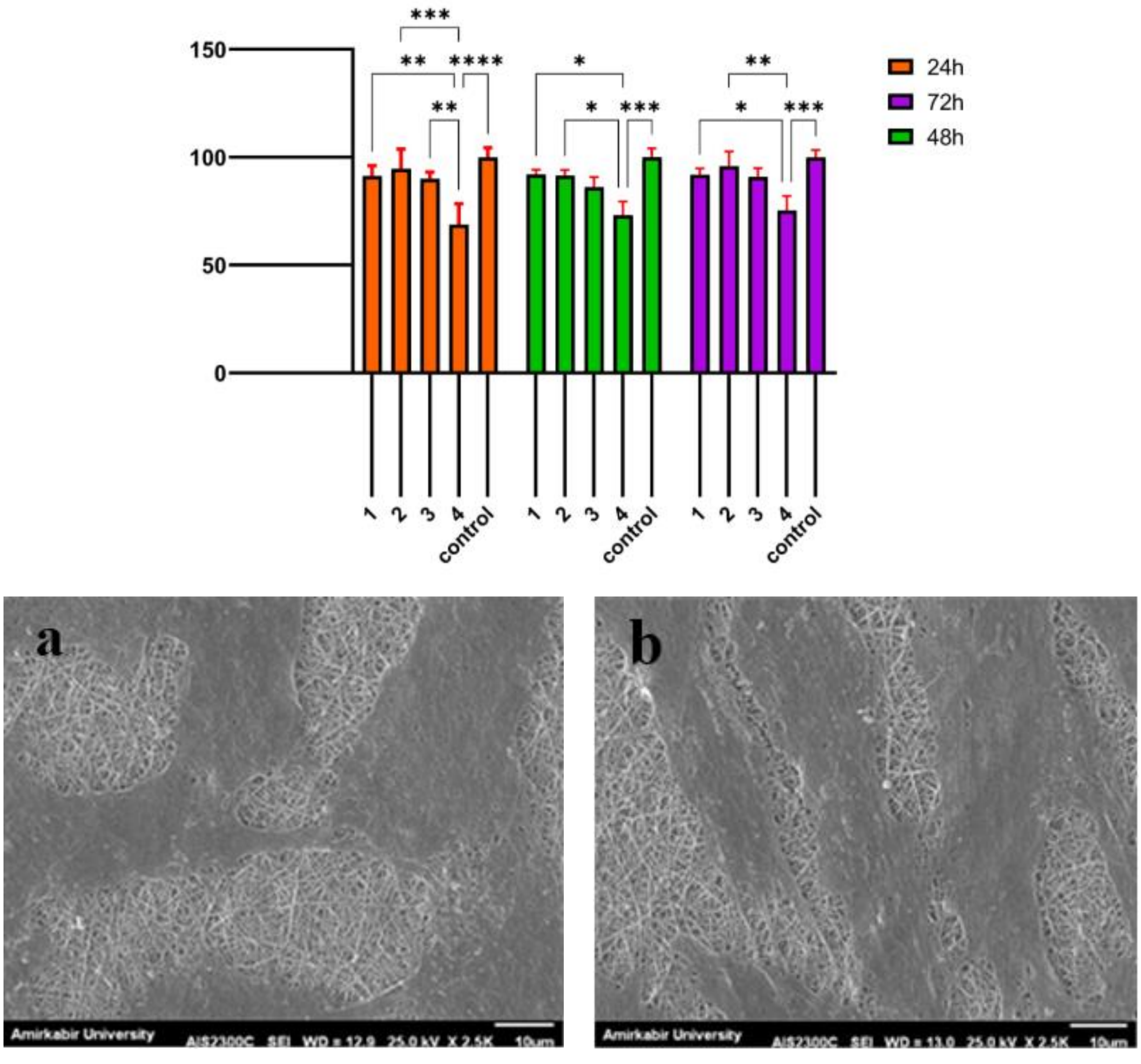

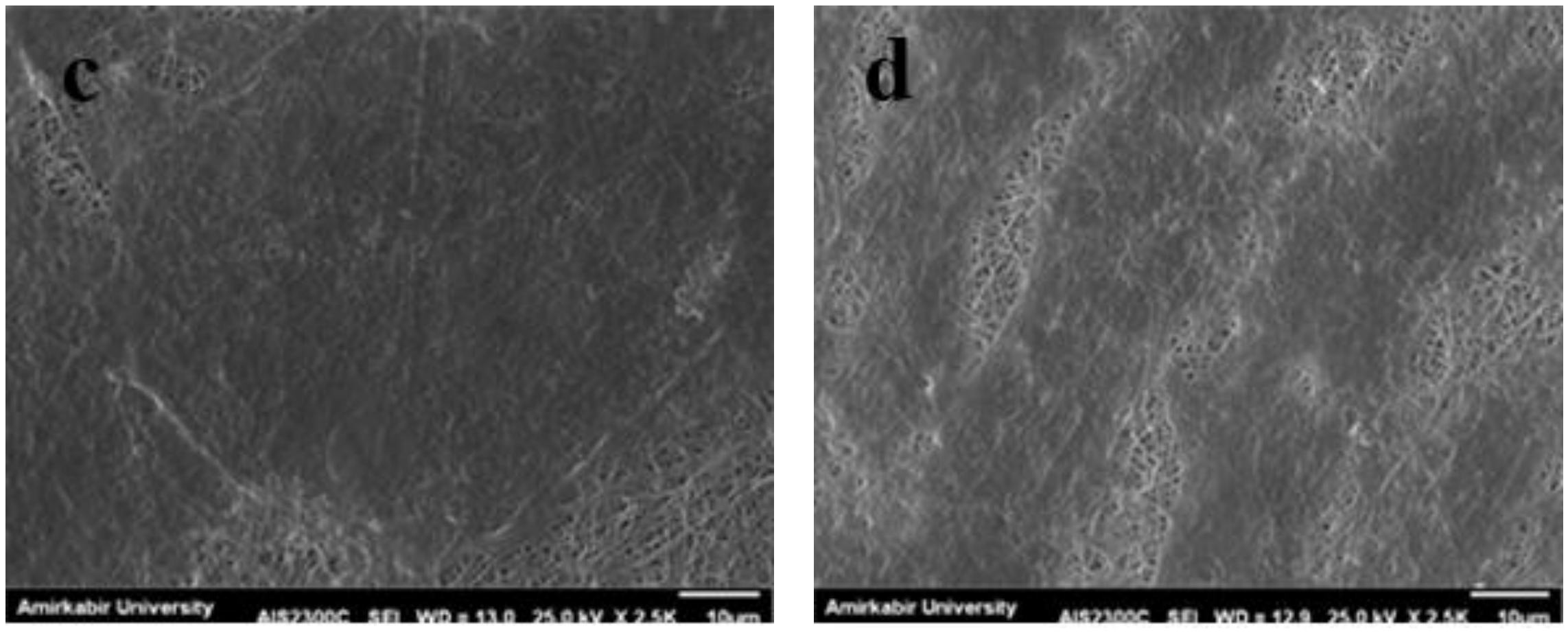

**Figure 14.** A) Cell viability assay of bilayer of PXDDA/Gel/PCL nanofibers including capsule in different concentrations of 1) CD/N-ACNP (0.0005/0.001) g/mL, 2) CD/N-ACNP (0.001/0.001) g/mL, and 3) CD/N-ACNP (0.002/0.001) g/mL, 4) One layer of PXDDA/Gel/PCL nanofibers during 1, 2, 3 days (*$p < 0.05$, ** $p < 0.01$, *** $p < 0.001$, and ****$p < 0.0001$).
B) Cell attachment of a) One layer of PXDDA/Gel/PCL crosslinked nanofibers as a control sample, b) Bilayer of PXDDA/Gel/PCL nanofibers with coating capsule in concentration of b) CD/N-ACNP (0.0005/0.001) g/mL, b) CD/N-ACNP (0.001/0.001) g/mL, and c) CD/N-ACNP (0.002/0.001) g/mL between layer of them.

## 4. Conclusion

This study characterized the synthesis of poly(xylitol dodecanedioic acid) (PXDDA) polymer with a molecular weight of 4038 g/mol, derived from xylitol and dodecanedioic acid (DDA) monomers via polycondensation reaction. The results confirmed that PXDDA polymer and its nanofibers exhibit autofluorescence within the wavelength range of 420–720 nm. Additionally, the polymer demonstrated approximately 90% weight loss between 350–510°C, consistent with its amorphous structure. The main objective was the fabrication of nanofibers using PXDDA blended with gelatin and PCL polymer solutions, aimed at reducing nanofiber diameter by increasing polymer chain entanglement. Optimization using Response Surface Methodology (RSM) and Artificial Neural Network (ANN) models highlighted the Genetic Algorithm (GA) in the ANN framework, which predicted an optimal nanofiber diameter of 269 nm with a cost value of 0.0054. This prediction closely aligned with experimental results, where nanofibers with a diameter of 271 ± 70 nm were obtained. To enhance water absorption stability, GA/HCl vapor crosslinking was conducted in an oven at 35°C for 2.5 hours, increasing contact angle and controlling water absorption.

Additionally, N-ACNP served as a carrier for clindamycin (CD), an antibacterial agent, through capsule coatings between PXDDA/Gel/PCL crosslinked nanofiber layers. Drug encapsulation within N-ACNP pores further regulated drug release, ensuring controlled and sustained drug delivery. The findings demonstrated that higher drug concentrations in bilayer nanofibers contributed to increased contact angle, a higher swelling percentage in the optima point of capsule concentration, enhanced biodegradability, and improved antibacterial properties. Finally, cell viability and cell culture analysis revealed that CD in the capsule at an optimal concentration between nanofiber layers resulted in the highest cell viability over three days, alongside enhanced cell adhesion.